# Electric Field Control of Molecular Charge State in a Single-Component 2D Organic Nanoarray

*Dhaneesh Kumar*[†, §], *Cornelius Krull*[†, §], *Yuefeng Yin*[§, ±, †], *Nikhil V. Medhekar*[§, ±] *and Agustin Schiffrin*[†, §, *]

[†]School of Physics & Astronomy, Monash University, Clayton, Victoria 3800, Australia

[§]ARC Centre of Excellence in Future Low-Energy Electronics Technologies, Monash University, Clayton, Victoria 3800, Australia

[±]Department of Materials Science and Engineering, Monash University, Clayton, Victoria 3800, Australia

* agustin.schiffrin@monash.edu

ABSTRACT: Quantum dots (QD) with electric-field-controlled charge state are promising for electronics applications, *e.g.*, digital information storage, single-electron transistors and quantum computing. Inorganic QDs consisting of semiconductor nanostructures or heterostructures often offer limited control on size and composition distribution, as well as low potential for scalability and/or nanoscale miniaturization. Owing to their tunability and self-assembly capability, using organic molecules as building nano-units can allow for bottom-up synthesis of two-dimensional (2D) nanoarrays of QDs. However, 2D molecular self-assembly protocols are often applicable on metals surfaces, where electronic hybridization and Fermi level pinning can hinder electric-field control of the QD charge state. Here, we demonstrate the synthesis of a single-component self-assembled 2D array of molecules [9, 10-dicyanoanthracene (DCA)] that exhibit electric-field-controlled spatially periodic charging on a noble metal surface, Ag(111). The charge state of DCA can be altered (between neutral and negative), depending on its adsorption site, by the local electric field induced by a scanning tunneling microscope tip. Limited metal-molecule interactions result in an effective tunneling barrier between DCA and Ag(111) that enables electric-field-induced electron population of the lowest unoccupied molecular orbital (LUMO) and hence charging of the molecule. Subtle site-dependent variation of the molecular adsorption height translates into a significant spatial modulation of the molecular polarizability, dielectric constant and LUMO energy level alignment, giving rise to a spatially dependent effective molecule-surface tunneling barrier and likelihood of charging. This work offers potential for high-density 2D self-assembled nanoarrays of identical QDs whose charge states can be addressed individually with an electric field.

KEYWORDS: supramolecular self-assembly, scanning tunneling microscopy and spectroscopy, non-contact atomic force microscopy, low-dimensional nanostructures, double barrier tunneling junction, density functional theory, organic quantum dots.

The manipulation of charge in solid-state systems – at the single electron level and with nanoscale precision – allows for various potential applications,[1] *e.g.*, nanoelectronics, memory storage, classical and quantum computing,[2] and light-emitting devices.[3] Quantum dots (QD), with their atom-like discrete electronic states whose occupancy can be tuned by an external electric field, offer a viable platform for such nanoscale charge state control,[4] and the realization of such technologies can benefit from the design of well-defined, ordered QD arrays, where the charge state of each QD can be addressed individually and independently.[5] Conventional QDs consist of inorganic hybrid semiconductor structures that can be synthesized *via* wet chemistry approaches (colloidal nanodots),[6-8] thin film growth (*e.g.*, heterostructures),[9-12] and even bottom-up atomic manipulation[13] (*e.g.*, on-surface one-dimensional nanostructures). However, these systems can have broad size distributions[6-8] and often give rise to disordered ensembles (colloidal QDs), showing low potential for miniaturization (*e.g.*, in the case of heterostructures)[9-12] or scalability (*e.g.*, nanostructures).[13]

Alternatively, single organic molecules can act as QDs[14, 15] with controllable integer charge state (that is, orbital occupancy) and be used for applications in molecular-based information storage[16] and transistors.[17] This approach offers the capability to tune molecular electronic states, charging properties and chemical reactivity *via* a rational structural design. Importantly, it allows for the synthesis of ordered molecular QD nanoarrays on surfaces *via* supramolecular self-assembly, from the bottom-up. This can be used for realizing two-dimensional (2D) QD arrays on metal surfaces, where confinement of electronic surface states by organic and metal-organic frameworks give rise to localized wavefunctions with discrete energy levels.[18-20] The small size of these systems provides a pathway for QD array miniaturization down to the nanoscale, and for arrays scaling to meso and micrometer domains.[21] However, these QD systems rely on electronic states of the metal surface, and on-surface supramolecular self-assembly protocols are mostly applicable on metal substrates.[22, 23] In these scenarios, electric field charge state control is hindered due to screening of the applied field, pinning of the confined discrete surface states or molecular states to the metal Fermi level, and/or hybridization between molecular and surface electronic states.[24] By efficiently decoupling molecule and substrate electronic states – that is, creating an effective potential barrier that separates molecule and surface – it becomes possible to alter the charge state of the molecular adsorbate with an external electric field. Such decoupling potential barrier can be achieved by thin insulating films grown between molecule and surface. However, molecular self-assembly protocols on dielectrics are not as well-established as on metals and remain challenging.

Previously, field-controlled charging of single molecules was shown in hybrid self-assembled molecular islands on a metal, where interactions between two different molecular species result in an effective decoupling from the substrate.[25] Field-driven charging was also observed at the boundaries of single-component molecular islands on metals,[26] as well as for whole molecular islands behaving as single QDs.[27] Given the challenges in achieving optimal stoichiometry in hybrid molecular systems[25] and in controlling the size of molecular islands,[26, 27] these systems offer however limited potential for the realization of ordered, scalable, high density nanoarrays of QDs with individually addressable charge state. Although potentially beneficial for such purpose, nanoscale field-control of individual molecular charge state in ordered, single-component organic nanofilms has not been achieved yet.

Here, we report the atomic-scale morphology and the electronic properties of a self-assembled single-component molecular nanofilm consisting of hydrogen-bonded 9,10-dicyanoanthracene (DCA) molecules on Ag(111). By low-temperature scanning tunneling microscopy (STM) and spectroscopy (STS), we observed spatially periodic negative charging of individual DCA molecules induced by the local electric field at the STM junction, evidenced by Coulomb blockade at a single molecule level[25, 28] and indicative of an effective molecule-surface potential barrier. We used Density Functional Theory (DFT) to complement and interpret our experimental results. Our data shows that the susceptibility of charging of the molecule (that is, the minimum applied electric field required to alter its charge state from neutral to negative) varies periodically within the molecular nanofilm and commensurately with the underlying noble metal atomic lattice. We quantitatively explain this effect by small site-dependent variations of the molecule-surface distance measured by low-temperature tip-functionalized non-contact Atomic Force Microscopy (nc-AFM). These variations affect the molecules' polarizability, which in turn results in a spatially varying effective molecule-surface potential barrier width and zero-field energy level alignment of the lowest unoccupied molecular orbital ($V_{\mathrm{LUMO}}^{(0)}$). The spatial periodicity of the field magnitude required to charge a molecule defines a 2D superstructure of ordered organic QDs whose charge state can be addressed individually.

## RESULTS AND DISCUSSION

Figure 1a shows an nc-AFM image (tip functionalized with carbon monoxide;[29] see Methods) of DCA molecules within a self-assembled domain after deposition on Ag(111) (see Methods for sample preparation). We considered sub-monolayer molecular coverages that resulted in DCA domains larger than 200 × 200 nm$^2$. The molecules form a 2D monoclinic lattice with unit cell vectors $\vec{a}_1$ and $\vec{a}_2$ ($\|\vec{a}_1\| = 1.20 \pm 0.02$ nm; $\|\vec{a}_2\| = 0.99 \pm 0.01$ nm; $\angle(\vec{a}_1;\vec{a}_2) = 53 \pm 1°$). Like other aromatic molecules, DCA adopts a planar adsorption configuration on Ag(111).[30-33] The molecular axis along the anthracene group (yellow dashed line) follows a 5 ± 1° angle with respect to $\vec{a}_1$.The two-fold symmetric cyano groups – with their lone electron pairs – mediate in-plane directional hydrogen-bonding with the neighboring molecules' extremities of the anthracene group, resulting in non-covalent networking. Such non-covalent 2D molecular self-assembly based on in-plane hydrogen bonding involving cyano-containing molecules is well established.[34-36]

The STM image in Figure 1b shows a larger area of the DCA self-assembly at a bias voltage, $V_b$ = – 0.6 V. Each bright elliptical feature corresponds to a single DCA molecule (inset of Figure 1b). Similar to the nc-AFM image in Figure 1a, the DCA domain appears homogenous. Figures 1c – f correspond to STM images of the same area at different bias voltages ($V_b$ = – 2.2 V, – 2.4 V, – 2.6 V, – 3.0 V, respectively). We observe that some molecules show a change in contrast, with a significantly smaller apparent height (label '$M_0$' in inset of Figure 1c), that is, with a significantly lower conductance (since a smaller tip-sample distance is required to maintain the same tunneling current). This change in contrast depends on the applied bias voltage, with more molecules showing a smaller apparent height along $\vec{a}_1$ as the absolute value of $V_b$ increases. This change in imaging contrast is reversible; the molecule regains its normal appearance with decreasing absolute value of $V_b$. The molecules with altered contrast define a periodic superstructure, which, for this specific molecular domain, has unit cell vectors $\vec{c}_1 = 22 \cdot \vec{a}_1 - \vec{a}_2$ and $\vec{c}_2 = 6 \cdot \vec{a}_1 + 2 \cdot \vec{a}_2$ (Figure 1f), with $|\vec{c}_1| = 26.4 \pm 0.1$ nm, $|\vec{c}_2| = 7.5 \pm 0.1$ nm, and $\angle(\vec{c}_1; \vec{c}_2) = 11 \pm 1°$ (superstructure unit cell area: $37.8 \pm 4.0$ nm$^2$). Here, along $\vec{a}_1$, this bias-dependent imaging contrast behavior repeats after 50 molecules. Along this direction, we labeled the molecules sequentially as '$M_i$' ($i$ = 0, …, 49), with '$M_0$' corresponding to the molecule that shows a change in contrast at the smallest $|V_b|$.

Figure 2a corresponds to the Fourier Transform (FT) of the STM image in Figure 1b superimposed on the FT of an atomically resolved STM image of bare Ag(111) (reciprocal lattice unit cell vectors $\vec{v}_1{}^*$ and $\vec{v}_2{}^*$). We observe peaks related to the periodicity of the molecular self-assembly (red dashed circles; vectors $\vec{a}_1{}^*$ and $\vec{a}_2{}^*$) as well as lower frequency peaks (black triangles in Figure 2a inset; vectors $\vec{c}_1{}^*$ and $\vec{c}_2{}^*$). These lower frequency peaks correspond to a subtle real-space long-range modulation of the molecular domain STM apparent height in Figure 1b. Importantly, the Fourier space vectors $\vec{c}_1{}^*$ and $\vec{c}_2{}^*$ associated with this modulation correspond to the real space vectors $\vec{c}_1$ and $\vec{c}_2$ associated with the bias-dependent superstructure in Figures 1c-e. That is, the molecules (labelled '$M_0$') that have a slightly smaller (~0.05 Å) STM apparent height in Figure 1b (closer to the surface) are those that appear darker in Figure 1c as the bias voltage $V_b$ is lowered from positive to negative values. This can further be observed in the STM apparent height profiles and nc-AFM frequency shift curves $\Delta f(z)$ in Figures 2b, c (see Supporting Information; SI).

From the FT in Figure 2a, we deduced that for the molecular domain in Figures 1b – f, $\begin{pmatrix} \vec{a}_1 \\ \vec{a}_2 \end{pmatrix} = \begin{bmatrix} 199/50 & 0 \\ 14/25 & 3 \end{bmatrix} \begin{pmatrix} \vec{v}_1 \\ \vec{v}_2 \end{pmatrix}$, where $\vec{v}_1$ and $\vec{v}_2$ define a real-space unit cell of the Ag(111) atomic lattice, and therefore $\begin{pmatrix} \vec{c}_1 \\ \vec{c}_2 \end{pmatrix} = \begin{bmatrix} 87 & -3 \\ 25 & 6 \end{bmatrix} \begin{pmatrix} \vec{v}_1 \\ \vec{v}_2 \end{pmatrix}$. The DCA self-assembly is commensurate with the noble metal atomic lattice, with the molecular row along $\vec{a}_1$ parallel to one of the Ag(111) <1,1,0> crystallographic axes. The superstructure with unit cell vectors $\vec{c}_1$ and $\vec{c}_2$ is a Moiré structure resulting from the lattice mismatch[37-39] between the molecular domain and Ag(111). Therefore, the change in molecular STM imaging contrast observed at negative biases (Figures 1c-f) depends on the molecules' adsorption site; two DCA molecules with the exact adsorption site will exhibit the same bias-dependent change in STM imaging contrast. It is important to note that the long range of the STM contrast modulation periodicity along $\vec{a}_1$ results in molecular domains where the relationships between vectors $\{\vec{c}_1, \vec{c}_2\}$ and $\{\vec{v}_1, \vec{v}_2\}$ vary slightly but still remain commensurate (see SI).

To gain insight into the bias-dependent change in STM imaging contrast, we performed differential conductance ($\mathrm{d}I/\mathrm{d}V$) STS measurements at different locations within a molecular domain (Figure 3; note that this is a different domain from that of Figures 1 and 2). In Figure 3b, the $\mathrm{d}I/\mathrm{d}V$ spectrum taken on top of molecule $M_0$ (molecule that changes its STM contrast at negative bias with the smallest absolute value $|V_b|$; *e.g.,* inset Figure 1c) shows a dip at a voltage $V^* \approx$ -2.1 V. This dip in the $\mathrm{d}I/\mathrm{d}V$ spectrum is characteristic of Coulomb blockade, indicating surface-to-molecule electron transfer and negative charging of the molecule induced by the bias voltage at the STM junction.[25] The ring observed in the negative-bias d$I$/d$V$ map in Figure 3b (inset) corroborates this charging effect.[26, 40-44] The charging bias onset $V^*$corresponds to the bias below which the contrast of the molecule changes in STM imaging, with a very significant bias-dependent reduction of the apparent height (*e.g.,* dark molecules in Figures 1c-f). This contrast change is the result of bias-induced negative charging of DCA.

The charging onset $V^*$ depends on the location of the molecule within the molecular domain. Molecule $M_0$ exhibits the smallest absolute value $|V^*|$; the further a molecule is from $M_0$ along $\vec{a}_1$, the larger $|V^*|$ (see SI). For example, $V^* \approx -2.23$ V for molecule labelled $M_4$ along $\vec{a}_1$ in Figure 3a. Within the considered bias voltage window (*i.e.*, for $V_\mathrm{b} > \sim - 3.2$ V, since $V_\mathrm{b} < \sim 3.2$ V tends to damage the molecules in the nanofilm), we did not observe charging for molecules adjacent to $M_0$ in the next parallel molecular row (for example, molecule $M_{29}$).

At positive bias voltages, a d$I$/d$V$ spectrum taken on top of a molecule (Figure 3b) shows a sharp step-like feature with an onset at ~120 mV. We associate this feature with a 2D electron gas (2DEG) at the molecule-metal interface.[25, 45-47] Investigation of this interface 2DEG is beyond the scope of this work. When acquired in between two molecules along $\vec{a}_1$, a d$I$/d$V$ spectrum shows – in addition to the 2DEG step-like feature – a peak at an energy which depends on the location within the long-rage superstructure;  next to molecule $M_4$, the peak is located at ~0.34 V; next to $M_{29}$, at ~0.39 V (Figures 3a, b).

The d$I$/d$V$ maps taken in the vicinity of molecule $M_0$ at $V_\mathrm{b}$ = 0.15 and 0.30 V (Figures 3c-e) show features centered on the molecules (red arrows) that we attribute to topography (see SI). At $V_\mathrm{b}$ = 0.30 V, the map shows extra features located at the extremities of the anthracene groups, in between molecules along $\vec{a}_1$, associated with the ~335 mV peak (cyan arrows; Figures 3d, e). The features at the anthracene extremities are reminiscent of the DCA lowest unoccupied molecular orbital (LUMO) with its nodal planes along the cyano-cyano axis [see LUMO of DCA on Ag(111) calculated by density functional theory in Figure 3f]. We hence associate the d$I$/d$V$ peak at $V_b \approx$ 335 mV with the LUMO. This is consistent with previous work on DCA adsorbed on graphene/Ir(111).[48]

To further understand the bias-induced charging of DCA, we performed measurements of $V^*$ and of the energy position $V_{\mathrm{LUMO}}$ of the LUMO d$I$/d$V$ peak, as function of tip-molecule distance $z$, for different molecules ($M_0$ to $M_{39}$; here, the periodicity along $\vec{a}_1$ repeats after 39 molecules, that is, $M_0 = M_{39}$ and so on) along $\vec{a}_1$ (Figure 4a). Figure 4b shows plots of $V^*(z)$ and $V_{\mathrm{LUMO}}(z)$ for $M_0$, $M_5$ and $M_9$. The absolute value $|V^*|$ of the charging bias voltage increases linearly with increasing $z$ (*e.g.*, $V^* = -1.70$ V to –2.48 V for $\Delta z = 3.1$ Å, for $M_0$, that is, 46 % variation,). This linear relationship provides evidence that the negative charging of the molecule is driven by the electric field applied at the STM junction; the slope of $V^*(z)$ defines the gating electric field, $F_{\mathrm{g}}$, required to charge the molecule (*e.g.*, $F_{\mathrm{g}}$ = -2.48 ± 0.01 V/nm for $M_0$). The energy position $V_{\mathrm{LUMO}}(z)$ of the LUMO also decreases (slightly) as $z$ is increased (*e.g.*, $V_{\mathrm{LUMO}}$ = 0.300 to 0.294 V for $\Delta z$ =1 Å for $M_0$, that is, 2 % variation). This demonstrates that the molecular orbital energy level is not perfectly pinned to the Fermi level $E_F$ of Ag(111). Similar $z$-dependent variations of $V_{\mathrm{LUMO}}$ have been observed previously, but mostly on thin dielectric films.[49]

The observed field-induced charging and $z$-dependent measured $V_{\mathrm{LUMO}}$ of DCA is a strong indication that the interaction between molecule and surface is weak. Indeed, significant interactions would hamper charge localization[25, 49, 50] and would result in Fermi level pinning of the LUMO. We explain this limited metal-molecule interaction by the flat molecular adsorption, similar to other aromatic molecules,[30-33] and by the cyano group lone electron pairs mainly interacting with adjacent molecules' anthracene groups *via* in-plane directional hydrogen bonding.[35, 36] The molecule-surface system therefore behaves as if an effective potential barrier existed between DCA and Ag(111), similar to the case of molecules adsorbed on thin insulating film on metals.[49] Importantly, the proximity of the DCA LUMO to the Fermi level is a key factor enabling the field-induced on-metal charging of DCA. This is in contrast to DCA dimers [formed during deposition with Ag(111) held at low temperature] which exhibit charging at significantly larger electric field magnitudes due to a higher LUMO energy and a greater molecule-surface interaction *via* the cyano group lone electron pairs (see SI). It is important to note that field-induced charging of other cyano-functionalized aromatic molecules in ordered organic monolayers on metals has been observed previously.[25]

This allows us to model the STM junction as a double barrier tunneling junction (DBTJ),[25, 49, 51] where one potential barrier consists of the vacuum between tip and molecule (with a barrier width $z$), and the other barrier consists of the effective potential barrier between molecule and metal (barrier width $d_{\mathrm{eff}}$; Figure 4d). Within this model, when a negative bias voltage $V_{\mathrm{b}}$ is applied, the potential across the DCA-Ag(111) barrier drops, that is, the LUMO energy shifts towards $E_{\mathrm{F}}$. This potential drop and resulting LUMO energy shift depend linearly on the ratio between $d_{\mathrm{eff}}$ and total barrier width $(z + d_{\mathrm{eff}})$; they are null if $d_{\mathrm{eff}}/(z + d_{\mathrm{eff}}) = 0$. If the potential drop becomes equal to the intrinsic LUMO energy $V_{\mathrm{LUMO}}^{(0)}$ (that is, the LUMO energy with respect to $E_{\mathrm{F}}$ when $V_{\mathrm{b}} = 0$), the LUMO energy $V_{\mathrm{LUMO}}$ aligns with $E_{\mathrm{F}}$, that is, $V_{\mathrm{LUMO}} = 0$; an electron can then tunnel through the DCA-Ag(111) barrier and populate the LUMO, resulting in negative charging of the molecule.

When molecular charging occurs, $V_{\mathrm{b}} = V^*$. Within this model, the bias voltage at which the molecule charges is given by

$$V^*(z) = -\frac{V_{\mathrm{LUMO}}^{(0)}}{d_{\mathrm{eff}}} z - V_{\mathrm{LUMO}}^{(0)} \qquad (1),$$

with[25, 26] $F_{\mathrm{g}} = -\, V_{\mathrm{LUMO}}^{(0)} / d_{\mathrm{eff}}$, and the measured LUMO energy with respect to $E_{\mathrm{F}}$ is given by

$$V_{\mathrm{LUMO}}(z) = V_{\mathrm{LUMO}}^{(0)} \frac{d_{\mathrm{eff}}}{z} + V_{\mathrm{LUMO}}^{(0)} \qquad (2).$$

We determined $V_{\mathrm{LUMO}}^{(0)}$ and $d_{\mathrm{eff}}$ by fitting our measured $V^*(z)$ and $V_{\mathrm{LUMO}}(z)$ with Equations (1) and (2), respectively, for molecules $M_0$ to $M_{10}$ and $M_{25}$ to $M_{39}$ (Figures 4a, b; Methods). These equations also allowed us to estimate the $z = 0$ reference (same for all molecules; see Methods). Note that we did not observe field-induced charging (for $V_{\mathrm{b}}$ between $\sim -3.2$ and 0 V, *i.e.,* for a maximum applied electric field strength of $\sim$4.5 V/nm) and could not retrieve $V_{\mathrm{LUMO}}^{(0)}$ and $d_{\mathrm{eff}}$ (and hence $F_{\mathrm{g}}$) for $M_{11}$ to $M_{24}$, (Figure 4c); charging of these molecules is hindered due to Coulomb blockade at adjacent molecules, which charge at smaller $|F_{\mathrm{g}}|$ (see SI). It is important to note that, for a given molecule, the same values of $V_{\mathrm{LUMO}}^{(0)}$ and $d_{\mathrm{eff}}$ (combined with a general $z = 0$ reference) result in the best fitting of both $V^*(z)$ and $V_{\mathrm{LUMO}}(z)$. This demonstrates that the DBTJ model provides a good physical description of our system, with reliable retrieval of $V_{\mathrm{LUMO}}^{(0)}$ and $d_{\mathrm{eff}}$ for each molecule. Also, it provides compelling evidence that the LUMO is indeed the electronic state involved in the charging *via* Ag(111)-to-DCA electron transfer.

The retrieved values of $V_{\mathrm{LUMO}}^{(0)}$ and $d_{\mathrm{eff}}$ (and hence $F_{\mathrm{g}}$) vary for different molecules (Figure 4c) following the spatial periodicity of the molecular charging pattern (Figure 4a); that is, $V_{\mathrm{LUMO}}^{(0)}$ and $d_{\text{eff}}$ depend on the molecular adsorption site on Ag(111). For example, molecule $M_0$ exhibits the smallest $V_{\text{LUMO}}^{(0)} \approx 248$ mV and largest $d_{\text{eff}} \approx 1.0$ Å, resulting in the smallest absolute value of the gating field required for charging ($|F_{\text{g}}| \approx 2.48$ V/nm). For comparison, $V_{\text{LUMO}}^{(0)} \approx 276$ mV, $d_{\text{eff}} \approx 0.8$ Å and $|F_{\text{g}}| \approx 3.4$ V/nm for $M_9$ (Figure 4c). This spatially periodic variation of the susceptibility of charging for different molecules when exposed to an applied electric field is what gives rise to the charging superstructure observed in the STM images at a given bias voltage (Figure 1).

What is the physical mechanism behind the adsorption site dependence of $V_{\text{LUMO}}^{(0)}$ and $d_{\text{eff}}$? We explain the variation of $V_{\text{LUMO}}^{(0)}$ by screening effects from the underlying metal, that can stabilize an electron injected to the LUMO *via* an image charge and increase the molecular electron affinity by a stabilization energy[52, 53] $\Delta E(d)$:

$$V_{\text{LUMO}}^{(0)}(d) = V_{\text{LUMO}}^{(0)}(\text{d} \rightarrow \infty) - \Delta E(d) \qquad \text{with} \qquad \Delta E(d) = \frac{1}{4\pi\epsilon_0}\frac{e^2}{2d} \qquad (3),$$

where $d$ is the molecule-metal distance. Note that this classical approach is justified by the weak DCA-Ag(111) interaction observed and does not include quantum mechanical effects (*e.g.*, hybridization). Given the small spatially periodic variations of $d$ along $\vec{a}_1$ observed in nc-AFM and STM (that is, related to apparent heights measured at small bias absolute values, where molecules are neutral; Figures 2b, c), we can write $d(n) = d_{\text{min}} + \frac{\Delta d}{2}\left(1 - \cos\left(\frac{2\pi n}{\lambda}\right)\right)$, where $n$ is the molecular position ($n = 0$ corresponds to $M_0$) and $\lambda$ is the observed periodicity of $d$ along $\vec{a}_1$ for a specific molecular domain (here, for Figure 4a, $\lambda = 39$). By using this expression of $d(n)$ in Eq. (3), we fitted our measured $V_{\text{LUMO}}^{(0)}(d(n))$ (Figure 4c) and determined the parameters $d_{\text{min}}$ (minimum molecule-surface distance, corresponding to $M_0$) and $\Delta d$ (molecule-surface distance variation between $M_0$ and $M_{20}$). We found $d_{\mathrm{min}} \approx 2.8$ Å and $\Delta d \approx 0.1$ Å (see SI). These values are consistent with our experimental data (Figure 2b) and with DFT-based calculations of the DCA adsorption height (~2.85 Å; see SI). We also considered possible screening by the polarizable surrounding molecules; however, the corresponding increase in stabilization energy is at most ~15% of $\Delta E$ calculated for screening by the metal (see SI). This provides evidence that the observed spatially dependent variations of $V_{\mathrm{LUMO}}^{(0)}$ can be explained by very subtle differences in molecule-surface distance.

Whilst nc-AFM and STM suggest that the molecule-metal distance *d* is the smallest for $M_0$ (Figure 1), Figure 4c shows that $M_0$ exhibits the largest effective DCA-Ag(111) potential barrier width $d_{\mathrm{eff}}$. We explain this discrepancy by noting that $d$ and $d_{\mathrm{eff}}$ are related *via* $d_{\mathrm{eff}} = d/\epsilon_{\mathrm{r}}$, where $\epsilon_{\mathrm{r}}$ is the out-of-plane relative dielectric constant of the molecule-metal barrier.[25, 49] By considering $d_{\mathrm{eff}}$ (1.00 ± 0.02 Å and 0.81 ± 0.05 Å) and $d$ (2.80 Å and 2.85 Å) for $M_0$ and $M_{10}$, we found values for $\epsilon_{\mathrm{r}}$ of 2.80 ± 0.06 and 3.51 ± 0.22, respectively. These values are similar to those reported for NaCl/Au(111)[54] and NaCl/Ag(100),[55] and are consistent with the weak DCA-Ag(111) interaction observed and the fact that DCA behaves effectively on Ag(111) as on a thin insulator. For NaCl/Au(111), $\epsilon_{\mathrm{r}}$ increases with increasing film thickness due to size effects in the ultra thin material,[56, 57] consistent with DCA/Ag(111) where $\epsilon_{\mathrm{r}}$ increases with *d*. The relative dielectric constant $\epsilon_{\mathrm{r}}$ is a measure of the local out-of-plane polarizability of DCA/Ag(111). A larger $\epsilon_{\mathrm{r}}$ translates into a larger dipole moment induced by the applied field, screening the latter and resulting in a larger absolute value $|F_{\mathrm{g}}|$ of the gating field required to charge the molecule. The smaller value of *d* for $M_0$ results, at the same time, in a smaller $V_{\mathrm{LUMO}}^{(0)}$ and larger $d_{\mathrm{eff}}$ (due to a smaller $\epsilon_{\mathrm{r}}$); whereas an electron injected into the LUMO is easier to screen by the metal for $M_0$ than for $M_{10}$, it is not so for an external electric field applied perpendicular to the molecule-surface system. These effects are responsible for the substantial variations in the susceptibility of charging observed for the different molecules.

CONCLUSIONS

In summary, we have demonstrated the bottom-up synthesis of a self-assembled 2D nanoarray of organic molecules on a metal surface, whose charge states can be controlled individually by an applied electric field. Field-driven charging is enabled due to weak molecule-metal interactions. The gating field required to alter the molecular charge state depends strongly on the adsorption site on the surface. This is due to subtle variations of the molecule-surface distance, which result in site-dependent LUMO energies and effective molecule-surface potential barrier widths. The molecules of the array ($M_0$) that require the smallest absolute gating field strengths define a grid of decoupled nanoscale units whose charge state (0 or $-|e|$) can be addressed individually by a control field. Our study demonstrates the realization of dense and scalable self-assembled 2D arrays of organic QDs with charge states individually addressable *via* field effect, spanning > 1 $\mu m^2$ areas with QD densities of ~3 per 100 $nm^2$.

METHODS

**Sample preparation**

The DCA organic sub-monolayers were synthesized in ultrahigh vacuum (UHV) by deposition of DCA molecules (Tokyo Chemical Industry; > 95 % purity) from the gas phase onto a clean Ag(111) surface held at room temperature. The Ag(111) surface was cleaned by repeated cycles of $Ar^+$ sputtering and annealing at 790 K. The base pressure during molecular deposition was below $5 \times 10^{-10}$ mbar. DCA was sublimed at 373 K resulting in a deposition rate of ~150 $nm^2/s$.

The size of the considered DCA domains was limited by the area of the bare Ag(111) terraces and by the molecular coverage.

**STM & STS measurements**

All STM and STS measurements were performed at 4.3 K in UHV (~$1 \times 10^{-10}$ mbar) with an Ag-terminated Pt/Ir tip. All topographic STM images were taken in constant-current mode with the sample bias reported throughout the text. All STS measurements, unless otherwise stated, were performed by recording the tunneling current as a function of the tip-sample bias voltage in the junction. During these STS measurements, the tip-sample distance was stabilized with respect to a specified tunneling current set point at a fixed reference location. We then numerically differentiated the resulting $I$-$V$ data to obtain d$I$/d$V$ as a function of tip-sample bias voltage. We considered bias voltages $V_\mathrm{b}$ in the range from $\sim -3.2$ to $\sim +1.0$ V, since voltages with absolute values $|V_\mathrm{b}| > \sim 3.2$ V tend to damage the molecular nanofilm. Constant-height d$I$/d$V$ maps in Figures 3d, e were obtained by acquiring $I$-$V$ and computing d$I$/d$V$ curves pixel by pixel as a function of tip location. The d$I$/d$V$ map in Figure 3b inset was obtained with a lock-in technique by applying a small modulation with 7 mV amplitude and 1.13 kHz frequency to the bias voltage.

**Nc-AFM measurements**

All nc-AFM measurements were performed at 4.3 K in UHV (~$1 \times 10^{-10}$ mbar) with a qPlus sensor with an Ag-terminated Pt/Ir tip. Unless otherwise stated, we functionalized the tip with a carbon monoxide (CO) molecule by introducing CO gas into the UHV chamber ($5 \times 10^{-8}$ mbar for ~3 seconds) with the sample at 4.3 K and picking up a CO molecule on bare Ag(111) with the tip.[58] Nc-AFM imaging was performed at a constant the tip height stabilized with respect to a specified tunneling current set point at a fixed reference location. Nc-AFM $\Delta f(z)$ curve in Figure 2c was obtained by varying the tip-sample distance $z$ with respect to a tip height defined by a specified tunneling current set point.

**DBTJ model**

We obtained curves $V^*(z)$ and $V_{\text{LUMO}}(z)$ in Figure 4a by measuring $I$-$V$ data for negative (–0.02 to –3 V) and positive (–0.02 to 0.8 V) bias voltages, at different tip-sample distances $z$ (determined relative to a tunneling set point $V_b = -0.02$ V, $I_t = 5$ pA above $M_0$). We then calculated d$I$/d$V$ curves as above and determined $V^*$ at a given $z$ and given molecule as the energy position of the sharp dip at negative biases (see Figure 3b). We determined the energy position $V_{\text{LUMO}}(z)$ of the LUMO peak by fitting the d$I$/d$V$ curves at positive biases (see SI for more details). By combining Equations (1) and (2), we obtain (see SI):

$$\frac{\partial \log\left(V^*(z)\right)}{\partial z} * \left(\frac{V_{\text{LUMO}}(z)}{V^*(z)} - 1\right)\bigg|_{z=z_{\text{setpoint}}} = \frac{1}{z_{\text{setpoint}}} \qquad (4)$$

which allows us to determine the $z = 0$ reference. We then fitted $V^*(z)$ and $V_{\text{LUMO}}(z)$ with Equations (1) and (2), respectively, to determine parameters $V_{\text{LUMO}}^{(0)}$ and $d_{\text{eff}}$ for each molecule (see SI for more details).

**DFT calculations**

The electronic structures of isolated DCA and DCA adsorbed on Ag(111) (SI) were calculated by Density Functional Theory (DFT) as implemented in the Vienna *ab initio* simulation package (VASP).[59] The Perdew-Burke-Ernzerhof (PBE) form of the generalized gradient approximation (GGA) was used to describe electron exchange and correlation[60] while a semi-empirical functional (DFT-D2) was used to describe dispersion forces.[61] In all relevant calculations, the plane-wave kinetic energy cutoff was set to 600 eV and all structures were relaxed until the ionic forces were smaller than 0.01 eV/Å. The Brillouin zone was sampled in a 9 × 9 × 1 Γ-centered cell for accurate calculations of the electronic structures.

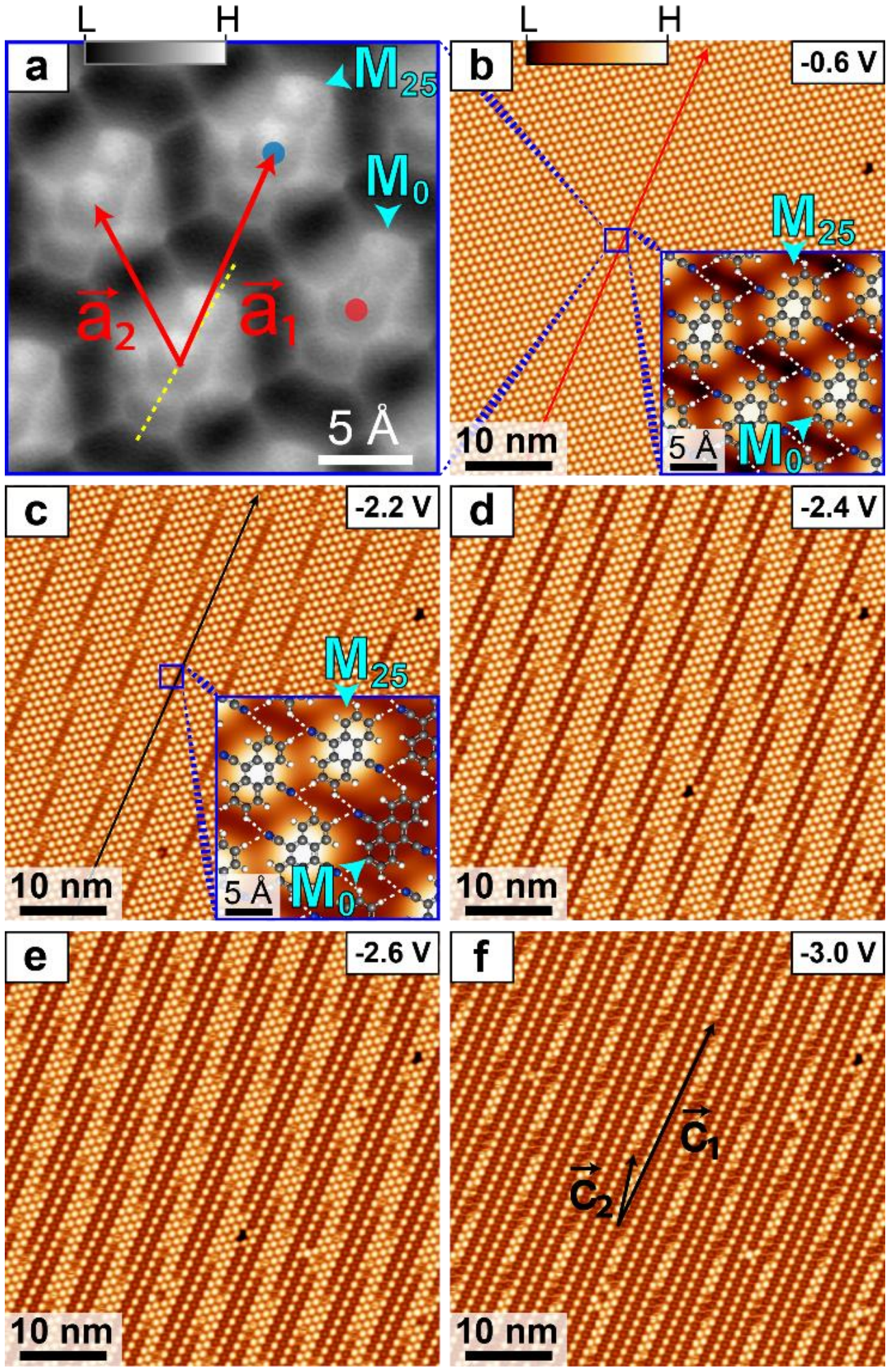

**Figure 1.** (a) Tip-functionalized nc-AFM image of DCA molecules in a self-assembled monolayer on Ag(111) (tip retracted 0.20 Å with respect to STM set point $V_b$ = –20 mV, $I_t$ = 200 pA, adjusted on top of molecule center). Long molecular axis (yellow dashed line) forms a ~5° angle with vector $\vec{a}_1$. (b) – (f) Constant-current STM imaging of DCA sub-monolayer on Ag(111) at $V_b$ = –0.6, –2.2, –2.4, –2.6, –3.0 V ($I_t$ = 50 pA). At $V_b$ = –0.6 V, the molecular domain appears homogenous; each bright ellipse is a DCA molecule. As $V_b$ decreases, some molecules are imaged darker, with a smaller apparent height [molecule labeled $M_0$ in (c) inset; $M_{25}$ indicates molecule with unchanged contrast], defining a periodic superstructure with unit cell vectors $\vec{c}_1$ and $\vec{c}_2$. The number of these dark molecules increases with increasing absolute value of $V_b$. Labels $M_0$ and $M_{25}$ in (a) and in insets of (b) and (c) indicate the same molecules.

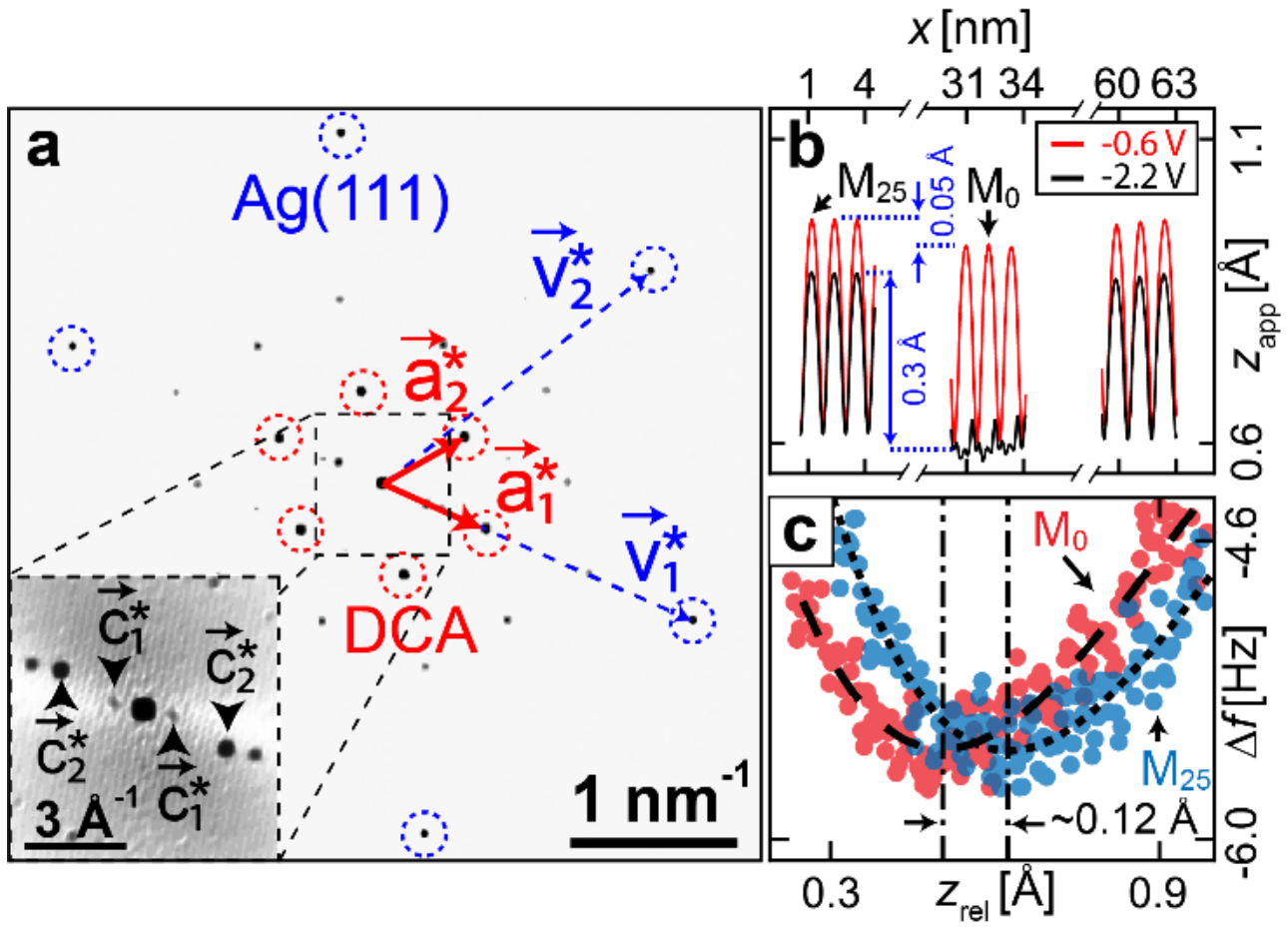

**Figure 2.** (a) Fourier transform (FT; $k$ = 1/wavelength) of STM image in Figure 1b, superimposed to the FT of an atomically resolved STM image of bare Ag(111). Vectors $\vec{v}_1{}^*$ and $\vec{v}_2{}^*$, $\vec{a}_1{}^*$ and $\vec{a}_2{}^*$, and $\vec{c}_1{}^*$ and $\vec{c}_2{}^*$ (indicated by black triangles in inset) define, respectively, unit cells of the reciprocal Ag(111) atomic lattice, of the reciprocal DCA domain lattice, and of the reciprocal superstructure lattice given by small variations of the molecular STM apparent height. These lattices are commensurate with each other. The reciprocal space vectors $\vec{c}_1{}^*$ and $\vec{c}_2{}^*$ correspond to the real space vectors $\vec{c}_1$ and $\vec{c}_2$ in Figure 1f. That is, the superstructure in Figures 1c – f given by bias-dependent altered contrast of DCA is perfectly correlated to the superstructure given by small apparent height variations in Figure 1b. (b) STM apparent height profiles of a molecular row along $\vec{a}_1$ at $V_b$ = – 0.6 V (red arrow in Figure 1b) and at $V_b$ = – 2.2 V (black arrow in Figure 1c). At $V_b$ = – 0.6 V, the apparent height of molecule $M_0$ is 0.05 ± 0.01 Å smaller than that of $M_{25}$. (c) Nc-AFM (tip functionalized with CO) resonance frequency shift $\Delta f$ as a function of relative tip-sample distance $z_{rel}$ ($z_{rel}$ = 0 corresponds to STM set point $V_b$ = –20 mV, $I_t$ = 200 pA measured on $M_0$), measured on $M_0$ (red) and $M_{25}$ (blue) (see Figure 1a). The turning point difference of ~0.12 Å confirms that $M_0$ adsorbs closer to the surface than $M_{25}$, and that molecule-surface distance depends on adsorption site, consistent with the STM apparent height variations.

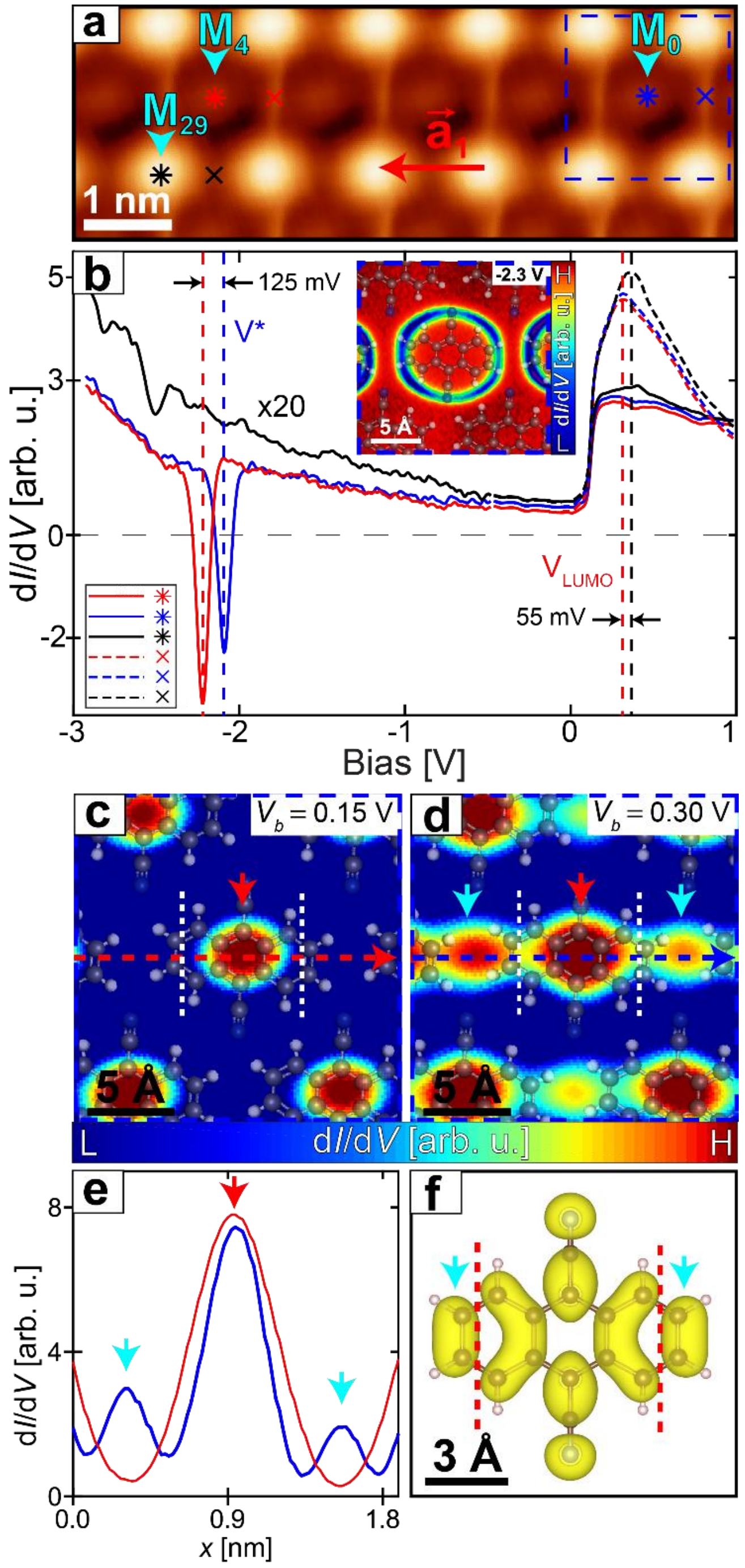

**Figure 3.** (a) Constant-current STM image of DCA domain ($V_b$ = -2.4 V, $I_t$ = 50 pA; corresponding to black rectangle in Figure 1e). At this bias voltage, molecules $M_0$ (blue dashed box) to $M_5$ show altered contrast, with a significantly reduced apparent height (dark imaging). Adjacent molecular row parallel to $\vec{a}_1$ shows unaltered apparent height (bright imaging). (b) d$I$/d$V$ spectra taken at locations indicated in (a) (set point $V_b = –0.5$ V, $I_t = 150$ pA). Spectra for dark molecules in (a) show negative differential conductance dip at negative bias voltages $V^*$, indicative of Coulomb blockade due to bias-induced negative charging of molecule. Negative voltage part of spectra was scaled by a factor of 20 for clarity. Spectra on top of molecules reveal a step-like feature with an onset at ~120 mV, indicative of a 2D interface state. Spectra in between molecules show a peak at positive bias voltages $V_{\text{LUMO}}$, associated with the lowest unoccupied molecular orbital (LUMO). Inset: constant-height d$I$/d$V$ map of molecule $M_0$ in (a), showing charging ring ($V_b = –2.3$ V; set point of $V_b = –0.5$ V, $I_t = 150$ pA on top of molecule). (c) – (d) Constant-height d$I$/d$V$ maps of molecule $M_0$ in (a), ($V_b = 0.15$ V and 0.3 V respectively; tip retracted 50 pm from set point $V_b$ = -20 mV, $I_t$ = 5pA on top of molecule). (e) d$I$/d$V$ profiles across maps in (c) (red), (d) (blue). Higher intensity at molecule center is due to topography (red arrows). For $V_b = 0.3$ V, higher intensity at anthracene extremities (cyan arrows) and nodal planes parallel to the axis defined by the cyano groups are characteristic of the LUMO (white dashed line). (f) DFT-calculated LUMO frontier of DCA in gas phase (isosurface level = $8.29 \times 10^{-4}$ e/Å$^3$).

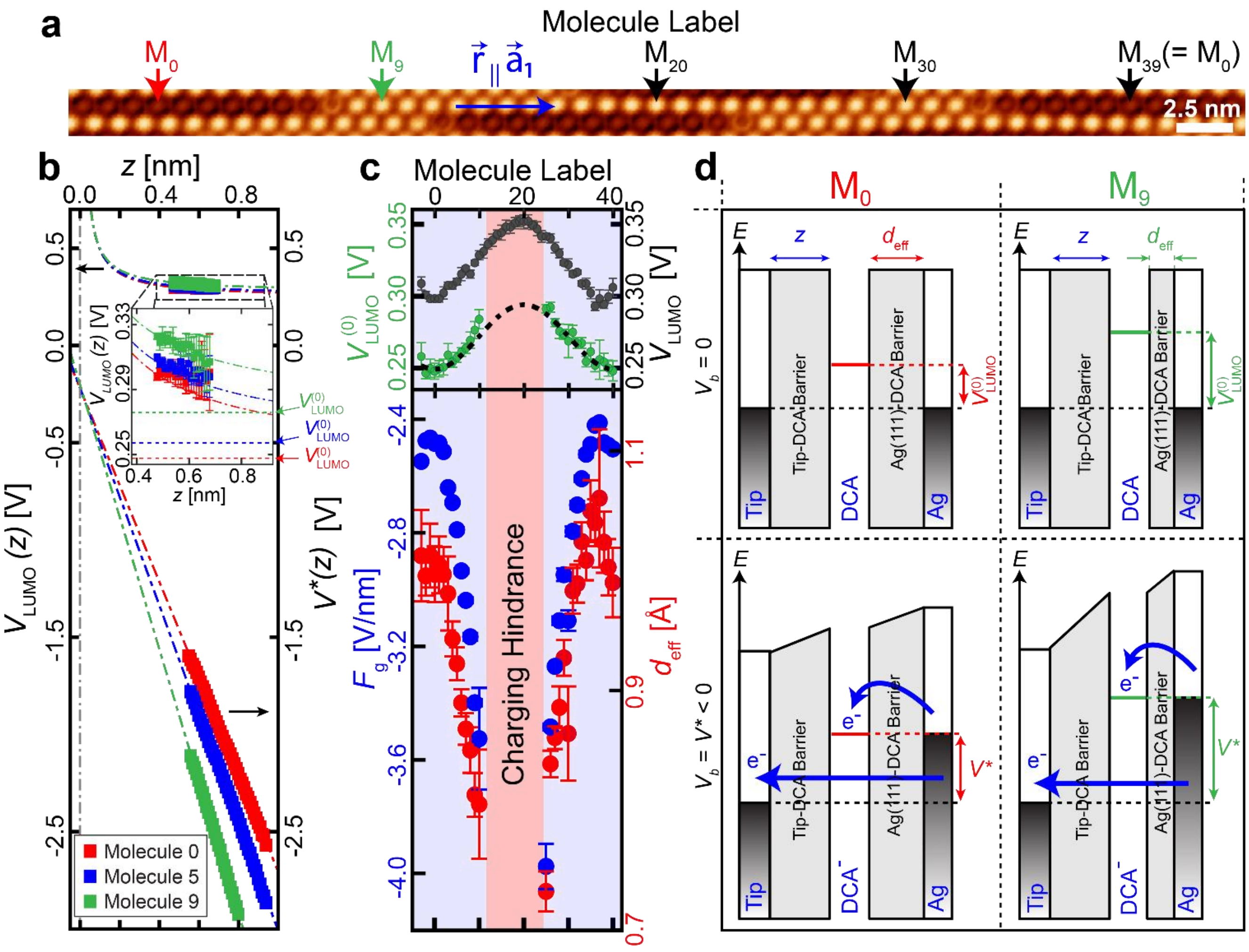

**Figure 4.** (a) STM image ($V_b$ = –2.4 V, $I_t$ = 50 pA) of molecular domain where the bias-dependent charging behavior of DCA along $\vec{a}_1$ repeats after 39 molecules. (b) Measurements of $V^*$ (bottom) and $V_{\mathrm{LUMO}}$ (top) for molecules $M_0$, $M_5$ and $M_9$ [labelled in (a)] as a function of tip-molecule distance, $z$. Dashed lines are fits of data based on the DBTJ model (Methods). Inset: zoomed-in plot of $V_{\mathrm{LUMO}}(z)$, asymptotically approaching $V_{\mathrm{LUMO}}^{(0)}$ as $z \to \infty$ ($z$ = 0.58 nm corresponds to STM set point $V_b$ = –20 mV, $I_t$ = 5 pA on top of $M_0$; actual tip height was derived from DBTJ model fit). (c) Top: variation of $V_{\mathrm{LUMO}}$ (black at $z$ = 0.47 nm) and $V_{\mathrm{LUMO}}^{(0)}$ (green) as a function of molecule label. Bottom: gating field, $F_{\mathrm{g}} = dV^*/dz = -\frac{V_{\mathrm{LUMO}}^{(0)}}{d_{\mathrm{eff}}}$, (blue) required to charge each molecule and effective molecule-metal distance, $d_{\mathrm{eff}}$ (red). Both $V_{\mathrm{LUMO}}^{(0)}$ and $d_{\mathrm{eff}}$ were extracted from DBTJ model fits of $V^*(z)$ and $V_{\mathrm{LUMO}}(z)$ data. For molecules $M_{11}$ to $M_{24}$, $V^*(z)$ could not be measured (and $F_{\mathrm{g}}$, $V_{\mathrm{LUMO}}^{(0)}$, $d_{\mathrm{eff}}$ could not be extracted) due to Coulomb blockade effects from neighbouring charged molecules. All error bars are ±1 standard deviation, except for $V^*(z)$ in (b) where they were omitted due to small size. (d) Schematic of DBTJ energy model depicting charging of molecules $M_0$ and $M_9$ [red and green arrows in (a)]. Charging bias onset $V^*$ corresponds to bias voltage $V_{\mathrm{b}}$ at which the substrate's Fermi level and molecule's LUMO align; $V^*$depends on $z$, $d_{\mathrm{eff}}$ and $V_{\mathrm{LUMO}}^{(0)}$. Top: $V_{\mathrm{b}} = 0$. Bottom; $V_{\mathrm{b}} = V^*$.

AUTHOR INFORMATION

**Corresponding Author**

* agustin.schiffrin@monash.edu, +61 3 9905 9265

ACKNOWLEDGMENT

D.K., Y.Y. and N.V.M. acknowledge funding support from the Australian Research Council (ARC) Centre of Excellence in Future Low-Energy Electronics Technologies. A.S. acknowledges funding support from the ARC Future Fellowship scheme (FT150100426). Y.Y. and N.V.M. acknowledge computational support from the Monash high performance computing cluster, the National Computing Infrastructure (NCI) and the Pawsey Supercomputing Facility.

TOC GRAPHIC

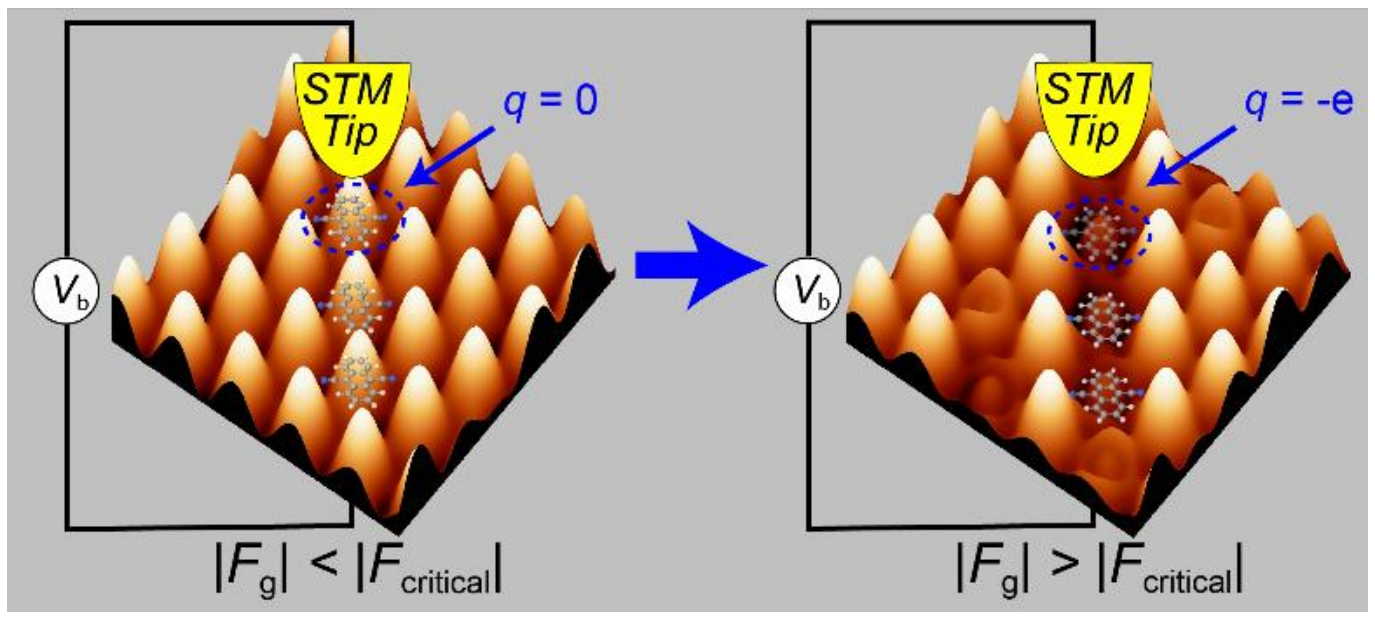

ASSOCIATED CONTENT

**Supporting Information**. The Supporting Information is available free of charge on the XX website at DOI: XX. Apparent height of DCA/Ag(111); DCA/Ag(111) domains and superstructure periodicity; Topography contribution to constant height d*I*/d*V* maps; Electric-field induced Ag(111)-to-DCA electron transfer; Charging hindrance: lateral Coulomb blockade; Double-barrier tunneling junction (DBTJ) model; Calculation of stabilization energy Δ*E*; DFT-calculated adsorption height; Formation of DCA dimers at low substrate temperature; Local contact potential difference (LCPD) measurements; Telegraph current signal of molecule with bistable charge state; d*I*/d*V* STS measurements for molecules $M_0$ to $M_{39}$ (negative bias range).

## Supporting Information

# Electric Field Control of Molecular Charge State in a Single-Component 2D Organic Nanoarray

*Dhaneesh Kumar*[†, §], *Cornelius Krull*[†, §], *Yuefeng Yin*[§, ±, †], *Nikhil V. Medhekar*[§, ±] *and Agustin Schiffrin*[†, §, *]

[†]School of Physics & Astronomy, Monash University, Clayton, Victoria 3800, Australia

[§]ARC Centre of Excellence in Future Low-Energy Electronics Technologies, Monash University, Clayton, Victoria 3800, Australia

[±]Department of Materials Science and Engineering, Monash University, Clayton, Victoria 3800, Australia

**Corresponding Author**

* agustin.schiffrin@monash.edu

## S1. Apparent height of DCA/Ag(111)

In order to evaluate the differences in adsorption height of different molecules, we measured the nc-AFM frequency shift $\Delta f$ as a function of tip-molecule distance $z$ above different molecules, using a CO-functionalized tip. The $z$ positions of the $\Delta f(z)$ minima allow us to estimate differences in actual tip-sample distance (and hence in actual molecule adsorption heights). For example, Figure S1b shows that the adsorption height of molecule $M_0$ is ~0.12 Å smaller than that of $M_{25}$. Table S1 shows height adsorption differences between $M_0$ and $M_{25}$ measured with different CO-functionalized tips, resulting in a weighted mean adsorption height difference of 0.08 ± 0.02 Å.

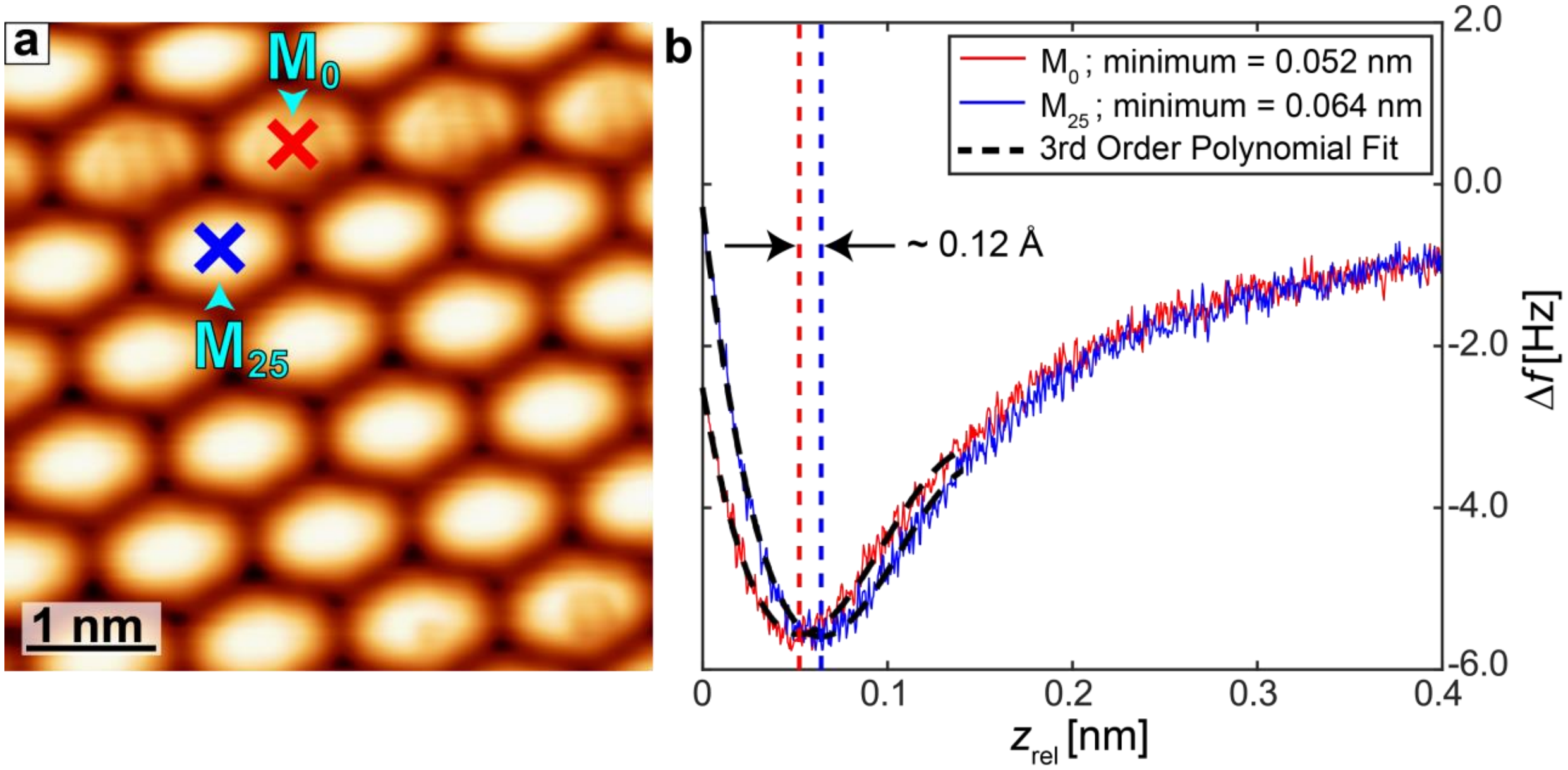

**Figure S1.** (a) STM image of DCA/Ag(111) acquired with a CO-functionalized tip ($V_b$ = -2.4 V, $I_t$ = 50 pA). Molecules marked with red and blue crosses correspond to molecules $M_0$ and $M_{25}$ respectively. (b) Nc-AFM frequency shift as a function of tip-molecule distance, $\Delta f(z)$, for $M_0$ (red) and $M_{25}$ (blue). The difference in $z$ positions of $\Delta f(z)$ minima between the two curves corresponds to the difference in adsorption height between the two molecules. The minima positions were determined by a $3^{rd}$ order polynomial fit (dashed black curve).

This (subtle) height variation can also be seen in nc-AFM imaging of the molecular domain (Figure S2a). Figure S2b shows the Fourier transform (FT) of the nc-AFM image in Figure S2a. Fourier-space peaks with vectors $\vec{c}_1^{\,*}$ and $\vec{c}_2^{\,*}$ indicate low-frequency modulation identical to that observed in STM imaging (Figure 2a of main text).

| Datasets | $\Delta\mathrm{h}_{ads} = \mathrm{h}_{min}^{(25)} - \mathrm{h}_{min}^{(0)}$ |
|---|---|
| Tip 1 | 0.024 Å |
| Tip 2 | 0.062 Å |
| Tip 3 | 0.117 Å |
| Tip 4 | 0.109 Å |
| Tip 5 | 0.089 Å |

**Table S1.** Difference in adsorption height, $\Delta\mathrm{h}_{ads}$, between molecules $M_0$ ($\mathrm{h}_{min}^{(0)}$: $z$ position of $\Delta f(z)$ minimum for $M_0$) and $M_{25}$ ($\mathrm{h}_{min}^{(25)}$), measured with 5 different CO-functionalized tips. This results in a mean difference in adsorption height, $\langle\Delta\mathrm{h}_{ads}\rangle$ = 0.08 ± 0.02 Å (the uncertainty reported is the standard error of the mean).

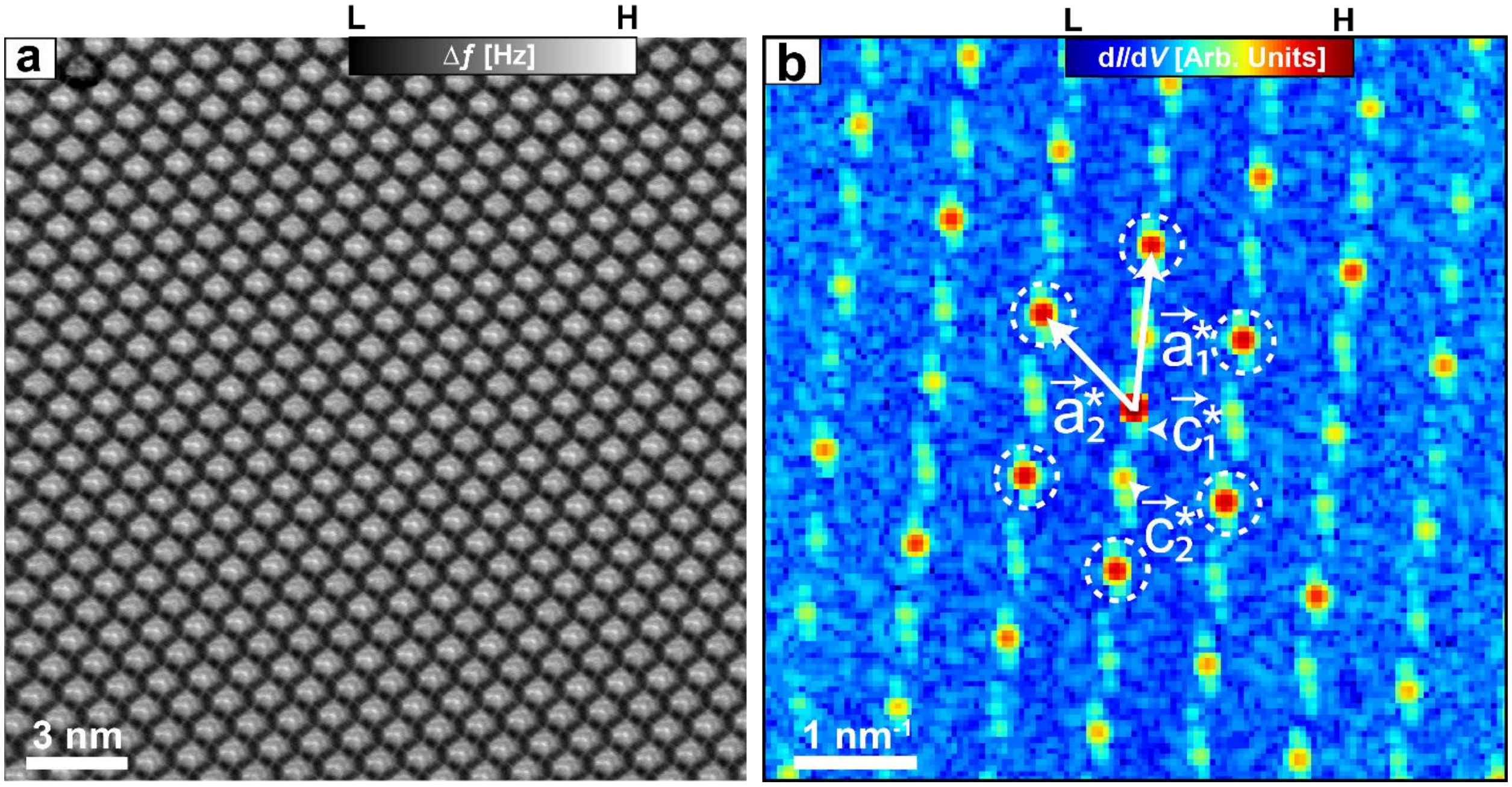

**Figure S2.** (a) Constant-height nc-AFM $\Delta f$ image of DCA/Ag(111) taken with CO-functionalized tip (tip height determined by STM set point $V_b$ = -20 mV, $I_t$ = 200 pA on a DCA molecule; imaging at $V_b = 0$ V). (b) FT of nc-AFM image in (a) showing peaks with vectors $\vec{a}_1{}^*$ and $\vec{a}_2{}^*$corresponding to the periodicity of the molecular lattice (dashed circles) and low frequency peaks with vectors $\vec{c}_1^*$ and $\vec{c}_2^*$ due to the subtle long-range modulation of the molecules' adsorption height (white triangles).

## S2. DCA/Ag(111) domains and superstructure periodicity

Given the three equivalent crystallographic axes <1,1,0> of Ag(111), and the relationship between these crystallographic axes and vectors $\{\vec{a}_1, \vec{a}_2\}$ defining a primitive unit of the molecular self-assembly (see main text), there are six equivalent molecular domains. That is, two mirror-symmetric domains associated with each crystallographic axis <1,1,0>; see Figure S3.

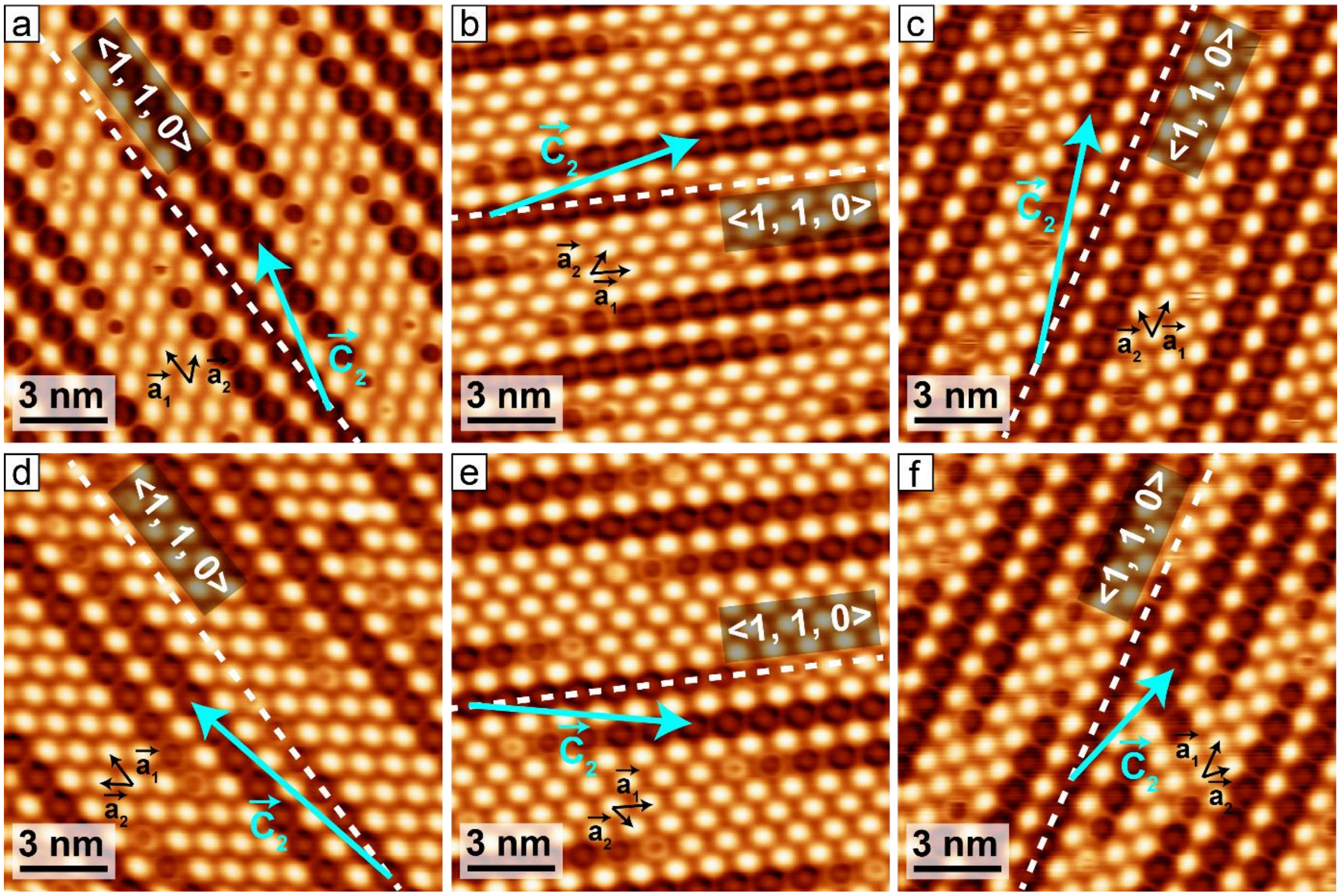

**Figure S3**. (a) – (f) Constant-current STM images ($V_b$ = -2.4 V, $I_t$ = 50 pA) showing the six equivalent molecular domains of DCA on Ag(111). White dashed lines: <1,1,0> crystallographic axis of Ag(111). Molecular arrangement in (a) [(b), (c)] is mirror-symmetric to that of (d) [(e), (f), respectively]. The molecular self-assembly unit cell vectors, $\{\vec{a}_1, \vec{a}_2\}$ (black arrows), and superstructure lattice vector, $\vec{c}_2$ (cyan arrows), for each domain are superimposed. Note that the superstructure for each of these six molecular domains differs relative to $\{\vec{a}_1, \vec{a}_2\}$ with (a) $\vec{c}_2 = 4\vec{a}_1 + 2\vec{a}_2$, (b) $\vec{c}_2 = 5\vec{a}_1 + 2\vec{a}_2$, (c) $\vec{c}_2 = 6\vec{a}_1 + 2\vec{a}_2$, (d) $\vec{c}_2 = 6\vec{a}_1 + 2\vec{a}_2$, (e) $\vec{c}_2 = 5\vec{a}_1 + 2\vec{a}_2$, and (f) $\vec{c}_2 = 3\vec{a}_1 + 2\vec{a}_2$.

It is important to note that the periodicity of the superstructure defined by the field-induced charging of DCA (unit cell vectors $\{\vec{c}_1, \vec{c}_2\}$; see main text) can change slightly throughout the extent of the molecular film. That is, the relationship between vectors $\{\vec{c}_1, \vec{c}_2\}$ and $\{\vec{a}_1, \vec{a}_2\}$ [and hence $\{\vec{v}_1, \vec{v}_2\}$ defining a unit cell of Ag(111)] can vary slightly from one molecular domain to another, but remains commensurate. In Figures 3 and 4 of the main text (where the periodicity of the superstructure along $\vec{a}_1$ is of 39 molecules), the relationship between $\{\vec{c}_1, \vec{c}_2\}$, $\{\vec{a}_1, \vec{a}_2\}$ is as follows:

$$\begin{pmatrix} \vec{c}_1 \\ \vec{c}_2 \end{pmatrix} = \begin{bmatrix} 33/2 & -1 \\ 6 & 2 \end{bmatrix} \begin{pmatrix} \vec{a}_1 \\ \vec{a}_2 \end{pmatrix}.$$

That is, $\vec{a}_1 = 155/39\,\vec{v}_1 (\approx 3.97\,\vec{v}_1)$, $\vec{a}_2 = 45/78\,\vec{v}_1 + 3\,\vec{v}_2 (\approx 0.577\,\vec{v}_1 + 3\,\vec{v}_2)$ and $\angle(\vec{a}_1; \vec{a}_2) = 50 \pm 2°$. This is very similar to the unit cell of the molecular domain in Figure 1 of the main text, where $\vec{a}_1 = 199/50\,\vec{v}_1 (= 3.98\,\vec{v}_1)$, $\vec{a}_2 = 14/25\,\vec{v}_1 + 3\,\vec{v}_2 (= 0.56\,\vec{v}_1 + 3\,\vec{v}_2)$ and $\angle(\vec{a}_1; \vec{a}_2) = 53 \pm 1°$. We can also observe this in Figure S3, where the vectors $\vec{c}_2$ for each of the six equivalent molecular domains are different relative to $\{\vec{a}_1, \vec{a}_2\}$. We attribute these slight variations to: (i) the long range of the superstructure periodicity with respect to the molecular arrangement (*i.e.*, the large number of molecular domain unit cells comprised in a superstructure unit cell can vary, with the relationship between vectors $\{\vec{a}_1, \vec{a}_2\}$ and $\{\vec{v}_1, \vec{v}_2\}$ remaining quasi-identical); and (ii) boundary effects, *e.g.*, confinement of molecular domains to Ag(111) terraces of finite size, and morphology of boundaries between different molecular domains. Even if the unit cell of the 2D charging lattice may vary slightly due to the long-range periodicity of the superstructure, these variations remain small, with the superstructure unit cell area – that is, the effective organic QD size – of $42 \pm 5$ nm$^2$.

## S3. Topography contribution to constant height d$I$/d$V$ maps

Figures S4a, b show the constant-height (d$I$/d$V$) ($x$, $y$) maps of Figures 3c, d in the main text, at $V_b$ = 0.15 and 0.30 V, after normalizing [*i.e.*, such that $0 \leq (dI/dV)_{norm}\ (x, y) \leq 1$] and dividing by the normalized tunneling current map $I_{norm}$ ($x$, $y$) ($0 \leq I_{norm}\ (x, y) \leq 1$). The increased d$I$/d$V$ intensity observed at the anthracene extremities when $V_b$ = 0.30 V (black arrows in Figure S4b and in the corresponding intensity line profile in Figure S4f) is a signature of the LUMO. This LUMO signature at the anthracene extremities is corroborated in Figures S4c-d, showing the normalized d$I$/d$V$ maps subtracted by the normalized tunneling current map $I_{norm}$ ($x$, $y$) (at $V_b$ = 0.15 and 0.30 V), and in Figure S4e, showing the difference $\Delta(dI/dV)_{\text{norm}}(x,y)$ between $(dI/dV)_{\text{norm}}(x,y)|_{V_{\text{b}}=300\text{ mV}}$ and $(dI/dV)_{\text{norm}}(x,y)|_{V_{\text{b}}=150\text{ mV}}$. The lack of intensity at the molecule center (red arrows) for $(dI/dV)_{\text{norm}}/I_{\text{norm}}$ (Figures S4a, b)), $(dI/dV)_{\text{norm}} - I_{\text{norm}}$ (Figures S4c, d) and $\Delta(dI/dV)_{\text{norm}}$ (Figure S4e) indicates that the features seen at the center of the molecules in Figures 3c, d of the main text are mainly an effect of topography.

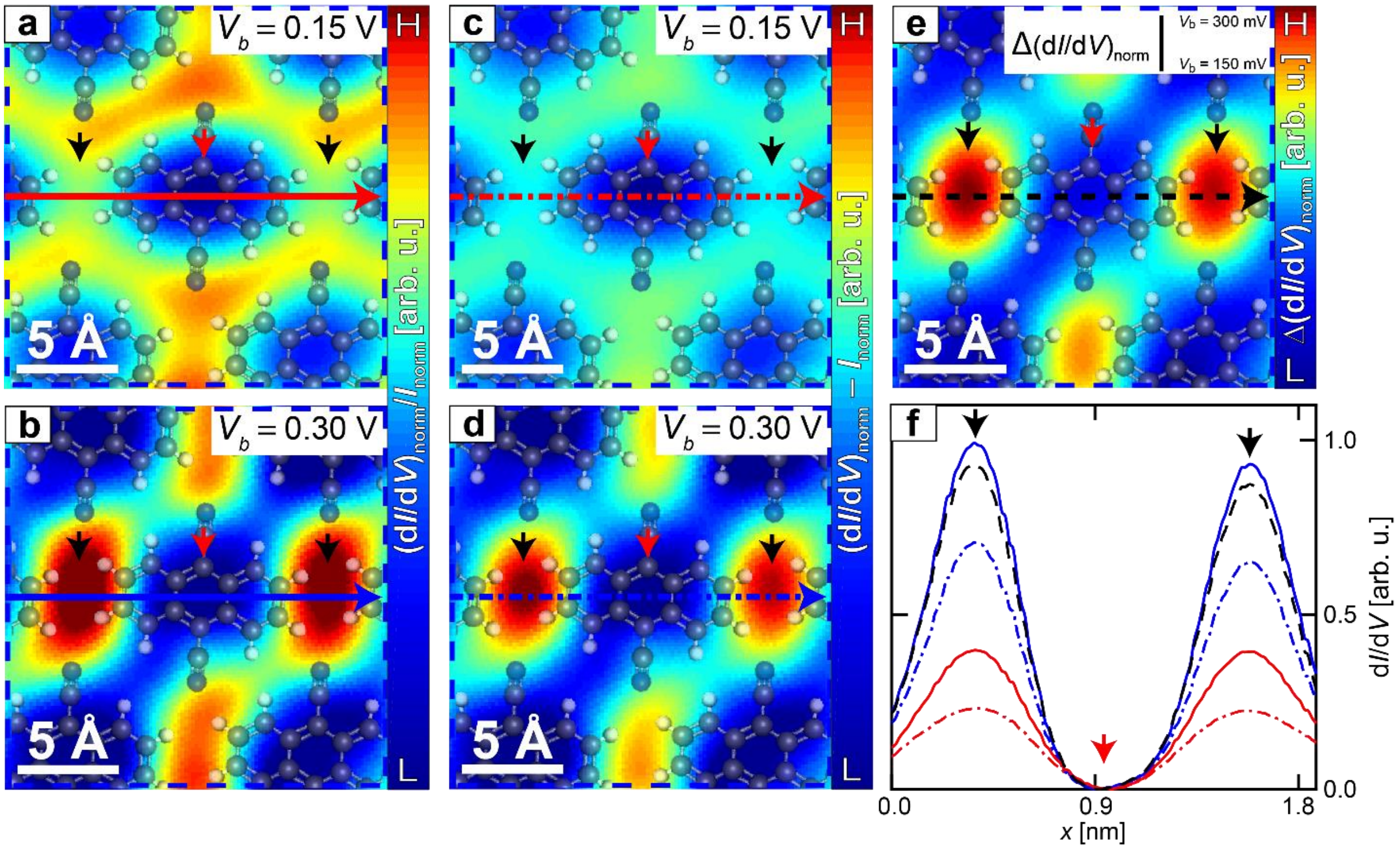

**Figure S4.** (a), (b) Constant-height $(dI/dV)_{\mathrm{norm}}/I_{\mathrm{norm}}$, (c), (d) $(dI/dV)_{\mathrm{norm}} - I_{\mathrm{norm}}$ (for bias voltages $V_{\mathrm{b}}$ displayed), and (e) $\Delta(dI/dV)_{\mathrm{norm}} = (dI/dV)_{\mathrm{norm}}|_{V_{\mathrm{b}}=300\,\mathrm{mV}} - (dI/dV)_{\mathrm{norm}}|_{V_{\mathrm{b}}=150\,\mathrm{mV}}$ maps, where $(dI/dV)_{\mathrm{norm}}(x,y) = (dI/dV)(x,y)/(dI/dV)_{\mathrm{max}}$ and $I_{\mathrm{norm}}(x,y) = I(x,y)/I_{\mathrm{max}}$, calculated using data in Figures 3c, d of main text. (f) Intensity line profiles along lines in (a) – (e). Increased intensity at black colored arrow locations when $V_{\mathrm{b}}$ = 300 mV are due to the LUMO which has strong contributions from the anthracene extremities (data taken with tip retracted 50 pm from STM set point $V_{\mathrm{b}} = -20\,\mathrm{mV}, I_{\mathrm{t}} = 5\,\mathrm{pA}$ measured on top of central molecule).

## S4. Electric-field-induced Ag(111)-to-DCA electron transfer

In the main text, we deduced that charging of DCA at negative bias voltages for $|V_{\mathrm{b}}| > |V^{*}|$ was the result of an electron transfer from Ag(111) to the LUMO. Figure S5a shows an STM image of DCA/Ag(111) at a bias voltage $V_{\mathrm{b}}$ = 0.33 V, above the LUMO onset (see main text), where DCA is neutral. Features in between the molecule centers, near the anthracene extremities, are characteristic of the LUMO (blue arrows in Figure S5a, c). At negative bias voltages, for $|V_{\mathrm{b}}| > |V^{*}|$, STM imaging of negatively charged molecules shows similar intermolecular features at the anthracene extremities (blue arrows in Figure S5b, d), identical to the neutral DCA LUMO. This similarity between imaging of neutral DCA unoccupied states (positive bias) and of negatively charged DCA occupied states (negative bias), with features characteristic of the LUMO, provides direct evidence that the field-induced charging of DCA is mediated by a population of the LUMO.

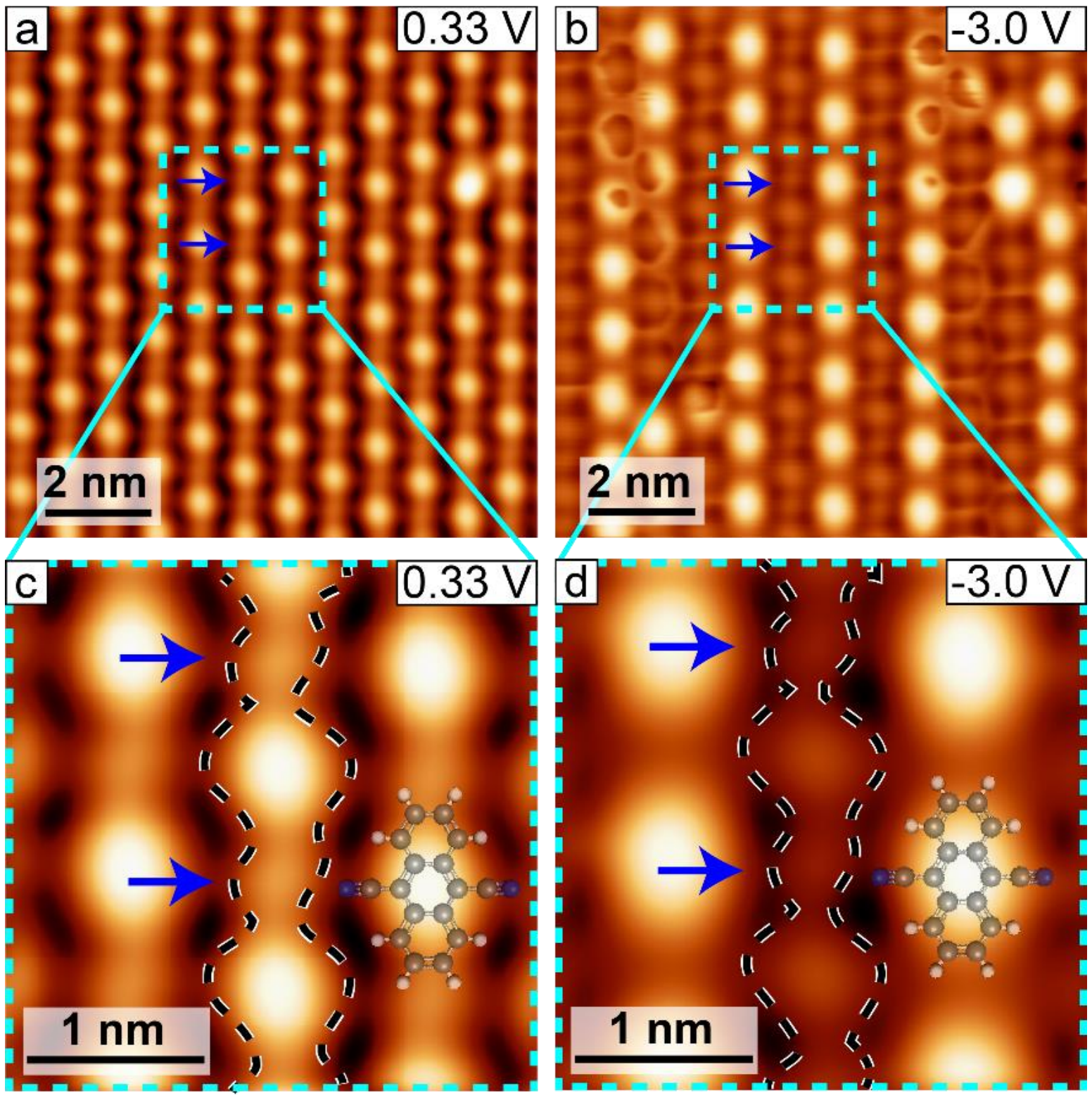

**Figure S5.** (a) – (b) Constant-current STM images of DCA monolayer on Ag(111) (a: $V_b$ = 330 mV, $I_t$ = 4 nA; b: $V_b$ = -3.0 V, $I_t$ = 300 pA). Nodal features (blue arrows) in-between DCA molecules associated with the LUMO can be seen for the neutral (a) and negatively charged species (b). (c) – (d) Zoomed-in STM images of (a) and (b) (cyan dashed square). Black dashed contours outline the same molecular row.

## S5. Charging hindrance: lateral Coulomb blockade

In Figure 4c of the main text, we could not measure $V^*$ for molecules $M_{11}$ to $M_{24}$ within the negative bias range ($V_\mathrm{b} > \sim -3.2$ V) and tip-sample distances considered. This is because, for this set of molecules $M_{11}$ - $M_{24}$, molecules in the adjacent row can become negatively charged at lower absolute field strengths, resulting in lateral Coulomb repulsion that increases the energy required for charging molecules $M_{11}$ - $M_{24}$ (*i.e.*, lateral Coulomb blockade). This is illustrated in Figure S6 for molecules $M_{11}$ and $M_{27}$ (different molecular domain from those in Figures 1-4 of main text; with a superstructure periodicity of 42 molecules along $\vec{a}_1$). Here, at $V_\mathrm{b} \approx$ -2.1 V, molecule $M_{27}$ charges while $M_{11}$ remains neutral. At $V_\mathrm{b}$ = -2.4 V (Figure S6b), $M_{27}$ can be charged if the lateral position of the STM tip is within the charging zone, defined by the charging ring around the molecule observed in the d$I$/d$V$ maps (see also inset of Figure 3b in main text). At higher absolute biases $|V_\mathrm{b}|$, the charging zone area increases (Figures S6c - e). Molecule $M_{11}$ charges at $V_\mathrm{b}$ = -2.8 V and at $V_\mathrm{b}$ = -3.0 V reverts to its neutral state. At this bias, the STM tip located above $M_{11}$ is within the charging zone of $M_{27}$; the negatively charging of adjacent $M_{27}$ results in lateral Coulomb repulsion that impedes charging of $M_{11}$ at this bias. At $V_\mathrm{b}$ = -3.17 V, this lateral Coulomb repulsion from charged $M_{27}$ is overcome and molecule $M_{11}$ charges again. Figure S6f shows a d$I$/d$V$ spectrum taken at $M_{11}$. The two dips at $V_\mathrm{b}$ = -2.8 and -3.17 V are associated with the two aforementioned charging events of $M_{11}$. The difference between the two dips' biases ($U \approx 0.37$ V) corresponds to the energy needed to overcome the lateral Coulomb field imposed by negatively charged $M_{27}$ and to charge $M_{11}$. Molecules $M_{11}$ to $M_{24}$ in Figure 4 of the main text are well within the charging zone of adjacent charged molecules (considering the range of biases and tip-sample distances used there), at bias voltages with absolute values smaller than $|V^*|$ required for charging. This means that absolute values of charging bias onsets for molecules $M_{11}$ to $M_{24}$ are increased by the charging energy, $U$, imposed by the adjacent charged molecules, putting them outside the bias range of our measurements (for all the tip-sample distances considered). Probing these molecules at higher bias absolute values proved difficult given the increased susceptibility of damaging the molecules within the nanofilm. We further note that, using the relative dielectric constant $\epsilon_{\mathrm{r}} = 3.51 \pm 0.22$ for $M_{10}$ in Figure 4 of the main text, $U_{\mathrm{Coulomb}} = (e^2/4\pi\epsilon_0\epsilon_{\mathrm{r}}) \times (1/r) = 0.41 \pm 0.03$ eV with the distance, $r \approx 1$ nm between $M_{11}$ and $M_{27}$ (see Figure S6e). This is consistent with our measured $U \approx 0.37$ V, corroborating our deduced $\epsilon_{\mathrm{r}}$ in the main text as well as our interpretation of the observed charging hindrance (*i.e.,* lateral Coulomb blockade).

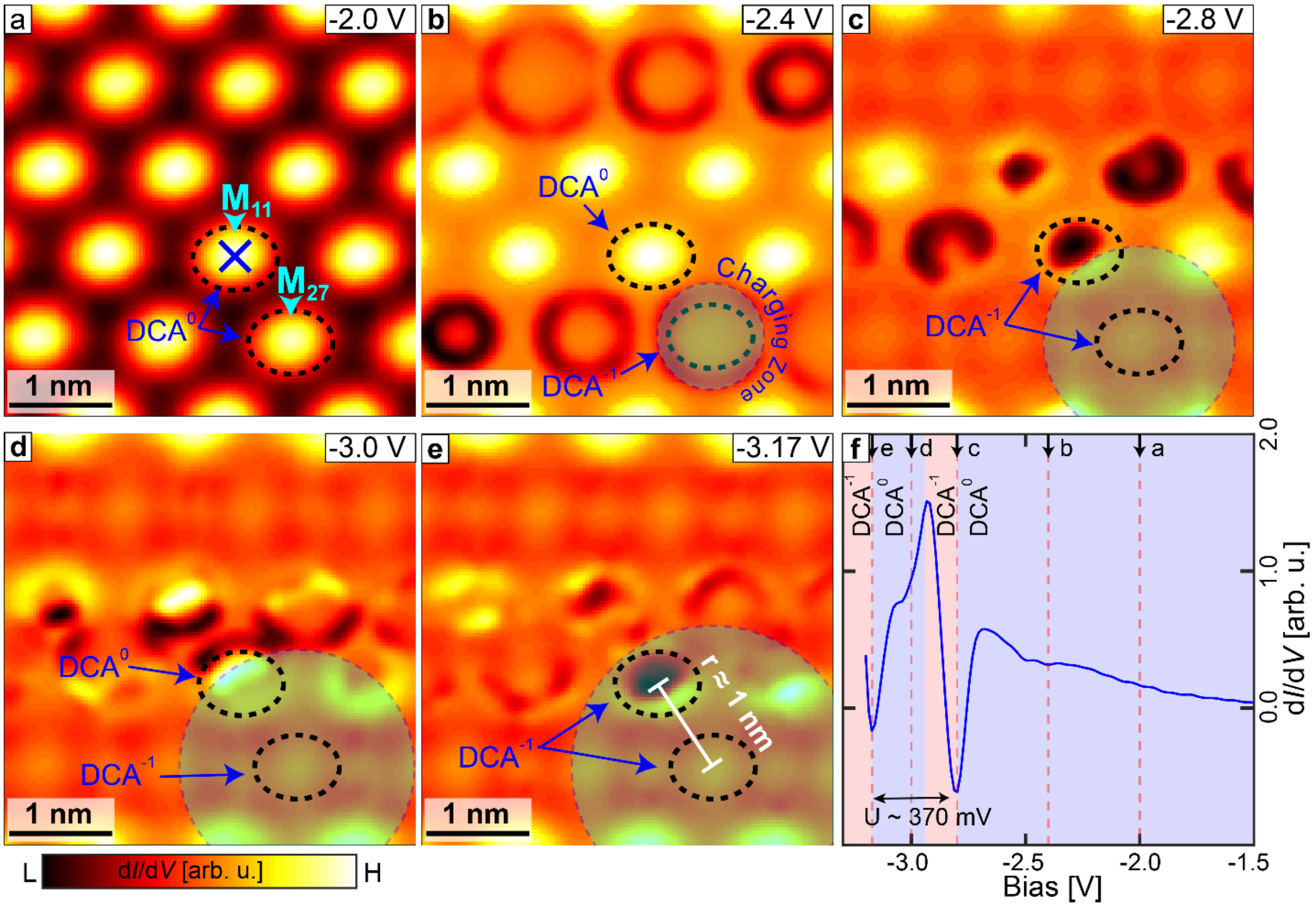

**Figure S6.** (a) – (e) Constant-height d*I*/d*V* maps (set point: $V_b$ = -0.02 V, $I_t$ = 5 pA measured on $M_{11}$) of DCA/Ag(111) for different negative bias voltages $V_b$ (shown on maps). For $V_b$ = -2 V (a), all DCA molecules are neutral ($DCA^0$). For $V_b$ = -2.4 V (b), $M_{27}$ is negatively charged ($DCA^{-1}$) and $M_{11}$ remains neutral. Shaded circular area marks the charging zone, defined by the maximum lateral distance between tip and molecule $M_{27}$ at which the latter can be charged, for a given bias voltage; the larger $|V_b|$ is, the larger the lateral tip-DCA distance can be for charging a specific molecule. For $V_b$ = -2.8 V (c), both $M_{11}$ and $M_{27}$ are charged. For $V_b$ = -3 V (d), the charging zone of $M_{27}$ overlaps with $M_{11}$ where now $M_{11}$: charging of $M_{11}$ is hindered by lateral Coulomb repulsion from charged $M_{27}$ and $M_{11}$ is neutral again. (e) For $V_b$ = -3.17 V, lateral Coulomb repulsion from charged $M_{27}$ is overcome and $M_{11}$ charges once more. (f) d*I*/d*V* curve taken on $M_{11}$ (set point: $V_b$ = -0.02 V, $I_t$ = 5 pA). Dips at $V_b$= -2.8 and -3.17 V are associated with the two charging events. The difference of $U \approx 0.37$ V between these two dips corresponds to the energy required to overcome the lateral Coulomb field from charged $M_{27}$. Blue (red) transparent background indicates when $M_{11}$ is neutral (charged).

## S6. Double-barrier tunneling junction (DBTJ) model

### Determination of $V_{\mathrm{LUMO}}$

We determined $V_{\mathrm{LUMO}}$ by fitting our d*I*/d*V* spectra at positive biases (Figure 3b of main text) with a sum of a Fermi-Dirac distribution-like function an exponentially modified Gaussian function:[1]

$$(\mathrm{d}I/\mathrm{d}V)_{\mathrm{fit}}(E) = \frac{a}{\exp(-(E - E_{\mathrm{IS}})/b) + 1} + \frac{c\exp\left(-\frac{1}{2}\left(\frac{E-g}{k}\right)^2\right)}{1 + \frac{(E-g)e}{k^2}}.$$

where *a*, *b*, *c*, *g*, *e*, *k*, are free fitting parameters. $E_{IS}$ is the interface state onset which was experimentally measured to be ~ 117 mV. We performed a constant background subtraction such that d*I*/d*V* = 0 for $V_{\mathrm{b}} < 0$ before the fitting procedure. We associate $V_{\mathrm{LUMO}}$ with the position (in energy) of the apex of the exponentially modified gaussian which is given by the following formula:

$$V_{\mathrm{LUMO}} = g - sgn(e)\sqrt{2}k\ \mathrm{erfcxinv}\left(\frac{|e|}{k}\sqrt{\frac{2}{\pi}}\right) + \frac{k^2}{e}\ ,$$

where erfcxinv(…) is the inverse of the scaled complementary error function. Figure S7a shows an example of a fitted d*I*/d*V* curve taken on $M_{20}$ (black circles: experimental data; solid red curve: fit) with $V_{\mathrm{LUMO}}$ = 355 ± 10 mV. The uncertainty in $V_{\mathrm{LUMO}}$ is determined through the propagation of the uncertainties (standard deviation) in the fitting parameters.

### Estimation of tip-molecule distance

We estimated the tip-molecule distance corresponding to the set point of $V_{\mathrm{b}}$ = -0.02 V, $I_{\mathrm{t}}$ = 5 pA at the location $M_0$, $z_{\mathrm{setpoint}}$, by using Equation (4) in the main text (Methods). Figure S7b shows the value obtained for each molecule (excluding $M_{11}$ to $M_{24}$) referred to in Figure 4 of the main text. The dashed blue line indicates the weighted mean value for all data shown, resulting in $\langle z_{\mathrm{setpoint}} \rangle = 5.8 \pm 0.3$ Å (uncertainty in weighted mean).

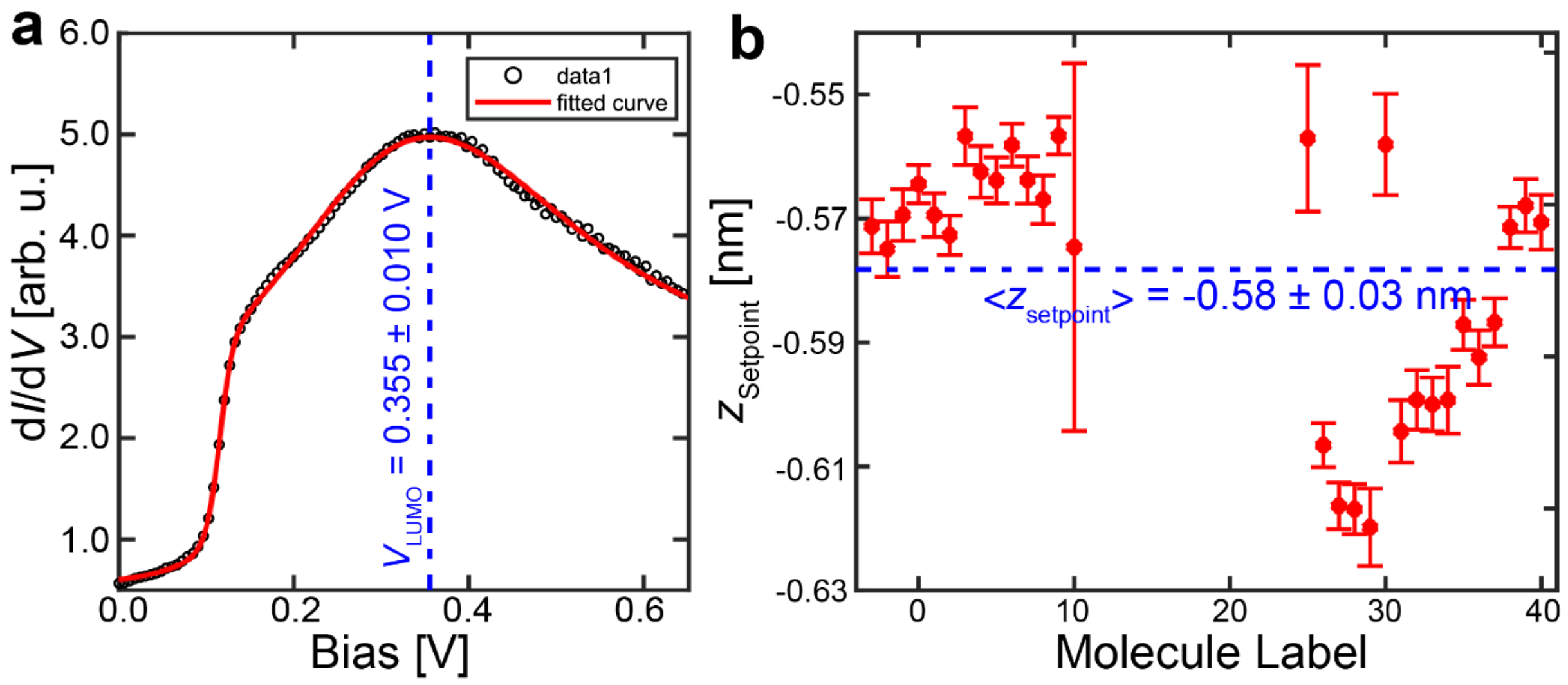

**Figure S7.** (a) Experimental d*I*/d*V* curve (black circles) and fit (solid red curve) taken above molecule M20. $V_{\mathrm{LUMO}}$ corresponds to the energy position of the exponentially modified Gaussian peak. (b) Estimated tip-molecule distance $z_{\mathrm{setpoint}}$ as a function of molecule label (corresponding to molecules in Figure 4 of main text). This results in $\langle z_{\mathrm{setpoint}} \rangle = -5.8 \pm 0.3$ Å which was used to fix *z* = 0 in Figure 4 of the main text.

## Determination of $V_{\mathrm{LUMO}}^{(0)}$ and $d_{\mathrm{eff}}$

For each molecule, we determined $V_{\mathrm{LUMO}}^{(0)}$ and $d_{\mathrm{eff}}$ by minimizing the following quantity $\chi$:

$$\chi = \sqrt{\chi_1^2 + \chi_2^2}$$

where:

$$\chi_1 = \frac{\sum_i \left( V^*(z_i) - V^*_{\mathrm{Eq}(1)}\left( V_{\mathrm{LUMO}}^{(0)}, d_{\mathrm{eff}}, z_i \right) \right)^2}{\left( \delta V^*(z_i) \right)^2},$$

and

$$\chi_2 = \frac{\sum_i \left(V_{\mathrm{LUMO}}(z_i) - V_{\mathrm{LUMO}}^{(\mathrm{Eq}(2))}\left(V_{\mathrm{LUMO}}^{(0)}, d_{\mathrm{eff}}, z_i\right)\right)^2}{\left(\delta V_{\mathrm{LUMO}}(z_i)\right)^2}.$$

The sum is performed over the tip heights $z_i$ considered. Functions $V_{\mathrm{Eq}(1)}^{*}$ and $V_{\mathrm{LUMO}}^{(\mathrm{Eq}(2))}$ are given by Equations (1) and (2) of the main text, respectively, whereas $V^{*}(z_i)$, $V_{\mathrm{LUMO}}(z_i)$, $\delta V^{*}(z_i)$, and $\delta V_{\mathrm{LUMO}}(z_i)$ are the measured $V^{*}(z)$ and $V_{\mathrm{LUMO}}(z)$ for each molecule and their respective uncertainties (at different heights).

## S7. Calculation of stabilization energy $\Delta E$

As mentioned in the main text, we determined the stabilization energy $\Delta E$ due to metal screening effects using Equation (3) (main text) where the metal-molecule distance $d$ was calculated using:

$$d(n) = d_{\mathrm{min}} + \frac{\Delta d}{2}\left(1 - \cos\left(\frac{2\pi n}{\lambda}\right)\right)$$

where $n$ is the molecule label (note that here we consider the same molecular domain as in Figure 4 of the main text), $\lambda = 39$ (corresponding to the superstructure periodicity along $\vec{a}_1$ in Figure 4 of the main text), $d_{\mathrm{min}}$ and $\Delta d$ are determined such that the following quantity $\chi_{\mathrm{d}}$ is minimized:

$$\chi_{\mathrm{d}} = \frac{\sum_i \left(V_{\mathrm{LUMO}}^{(0)}(n) - V_{\mathrm{LUMO}}^{(Eq(3))}(d_{\mathrm{min}}, \Delta d, n)\right)^2}{\left(\delta V_{\mathrm{LUMO}}^{(0)}(n)\right)^2}.$$

Figure S8a shows experimental data of $V_{\mathrm{LUMO}}^{(0)}$ (blue) and best fit according to this model (red), with $d_{\mathrm{min}} \approx 2.8$ Å and $\Delta d \approx 0.1$ Å. We further considered screening effects by the surrounding molecules, which can be polarized laterally (in plane) and increase the stabilization energy. Using the *Gaussian* software,[2] we calculated the molecular polarizability tensor $\boldsymbol{\alpha}$ of DCA in the gas phase (in CGS units):

$$\boldsymbol{\alpha}(\text{Å}^3) = \begin{pmatrix} \alpha_{xx} & \alpha_{xy} & \alpha_{xz} \\ \alpha_{yx} & \alpha_{yy} & \alpha_{yz} \\ \alpha_{zx} & \alpha_{zy} & \alpha_{zz} \end{pmatrix} = \begin{pmatrix} 41.40 & 0 & 0 \\ 0 & 35.82 & 0 \\ 0 & 0 & 8.17 \end{pmatrix}$$

where $x$ is along the long molecular axis (anthracene), $y$ is along the cyano-cyano axis and $z$ is perpendicular to the anthracene plane. Screening *via* classical electrostatic metal-molecule interaction changes the molecular in-plane polarizabilities $\alpha_{xx}$ and $\alpha_{yy}$ according to:[3]

$$\alpha_{xx,yy}{}' = \frac{\alpha_{xx,yy}}{1 - \frac{k\alpha_{xx,yy}}{(2d)^3}}$$

where $k = 1/(4\pi\epsilon_0)$ is the Coulomb constant and $d$ is the metal-molecule distance. The contribution to the stabilization energy of the *n*th molecule due to in-plane screening from the surrounding molecule then becomes:[4]

$$\Delta E^{(\mathrm{p})}(n) = \frac{k^2 e^2}{2} \sum_i \frac{\alpha_i'}{|\vec{R}_i|^4}$$

with

$$\alpha'_i = (\alpha_{xx}'\hat{\imath} + \alpha_{yy}'\hat{\jmath}) \cdot \hat{R}_i$$

where $e$ is the elementary electric charge, $\vec{R}_i$ is the vector displacement between the *n*th and *i*th molecule. Here, for a given molecule with label $M_n$, we consider in-plane screening from the six nearest neighbors with labels $M_{j_1}$ ... $M_{j_6}$, as indicated in Figures S8b. The nearest-neighbor metal-molecule distances $d_{j_1}$ ... $d_{j_6}$ for molecules $M_{j_1}$ ... $M_{j_6}$ depend on indexes $j_1 \ldots j_6$, which, for a given *n*, are determined by the periodicity of the charging superstructure as follows:

$$j_1 = n + 1 \bmod(39),$$

$$j_2 = n + 18 \bmod(39),$$

$$j_3 = n + 17 \bmod (39),$$

$$j_4 = n - 1 \bmod(39),$$

$$j_5 = n - 18 \bmod(39),$$

$$j_6 = n - 17 \bmod(39).$$

Using $d_{\mathrm{min}} \approx 2.8$ Å and $\Delta d \approx 0.1$ Å for the expression of $d(n)$ above, we found that $\Delta E^{(\mathrm{p})}(n)$ provides a further increase of ~15% to the stabilization energy for $M_{20}$ ($\Delta E^{(\mathrm{p})}(n = 20)/\Delta E(d) = \sim 0.15$) which in turn results in a reduction in $V_{\mathrm{LUMO}}^{(0)}$ as seen in Figure S8a (green curve). Since most of the stabilization energy is due to screening *via* molecule-metal interactions, we omitted these in-plane molecule-molecule interactions in the main text. Note that calculated molecular polarizabilities are often overestimated,[5-7] and therefore the actual contribution of in-plane molecule-molecule interactions to the total screening stabilization energy is likely smaller than our estimated ~15% (for $n = 20$) increase.

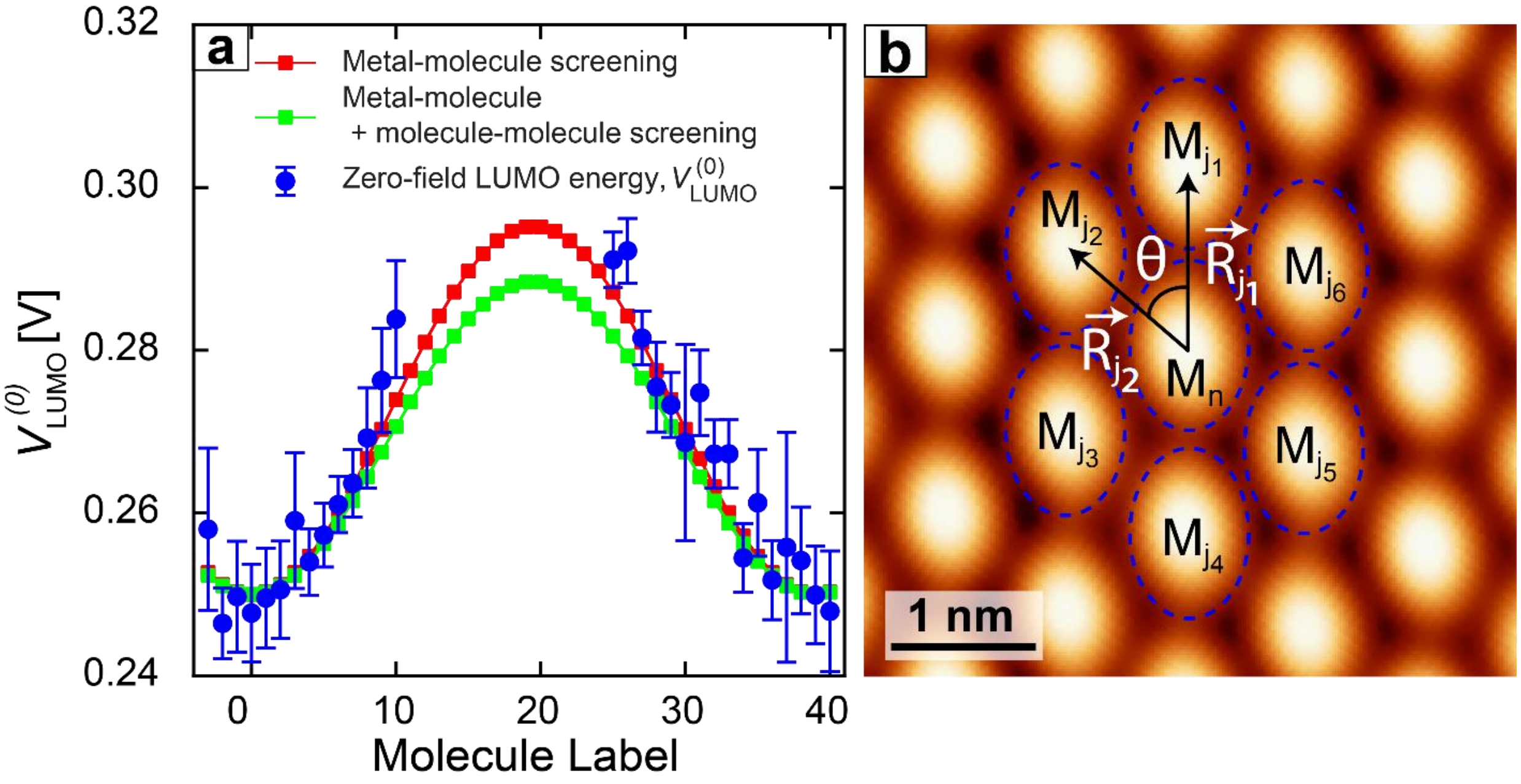

**Figure S8.** (a) Zero-field LUMO energy $V^{(0)}_{\mathrm{LUMO}}$ as a function of molecule label along $\vec{a}_1$. Blue circles: experimental data. Red: model fit with screening stabilization energy provided only by molecule-metal interactions. Green: model fit with screening stabilization energy provided by molecule-molecule and molecule-metal interactions. (b) Constant-current STM image of DCA molecule $M_n$ (here, $n$ = 0) surrounded by six nearest neighbors $M_{j_1} \dots M_{j_6}$ on Ag(111) ($V_b$ = -0.02 V, $I_t$ = 50 pA). In our model to calculate in-plane screening due to nearest-neighbor molecule-molecule interactions, we used $\|\vec{R}_{j_1}\|$ = 12.3 Å, $\|\vec{R}_{j_2}\|$ = 10.8 Å and $\theta$ = 50.2°, determined by experiment.

## S8. DFT-calculated adsorption height

We performed DFT calculations to determine the adsorption height of a single DCA molecule on Ag(111). For the DFT computation, we considered three inequivalent initial DCA adsorption sites and allowed the system to relax to a final energetically favorable configuration. Figure S9 shows the resulting adsorption heights, with slight variations depending on the adsorption site. This is consistent with our model assumptions. The average adsorption height is ~2.86 Å. Details of DFT computation can be found in the Methods section of the main text.

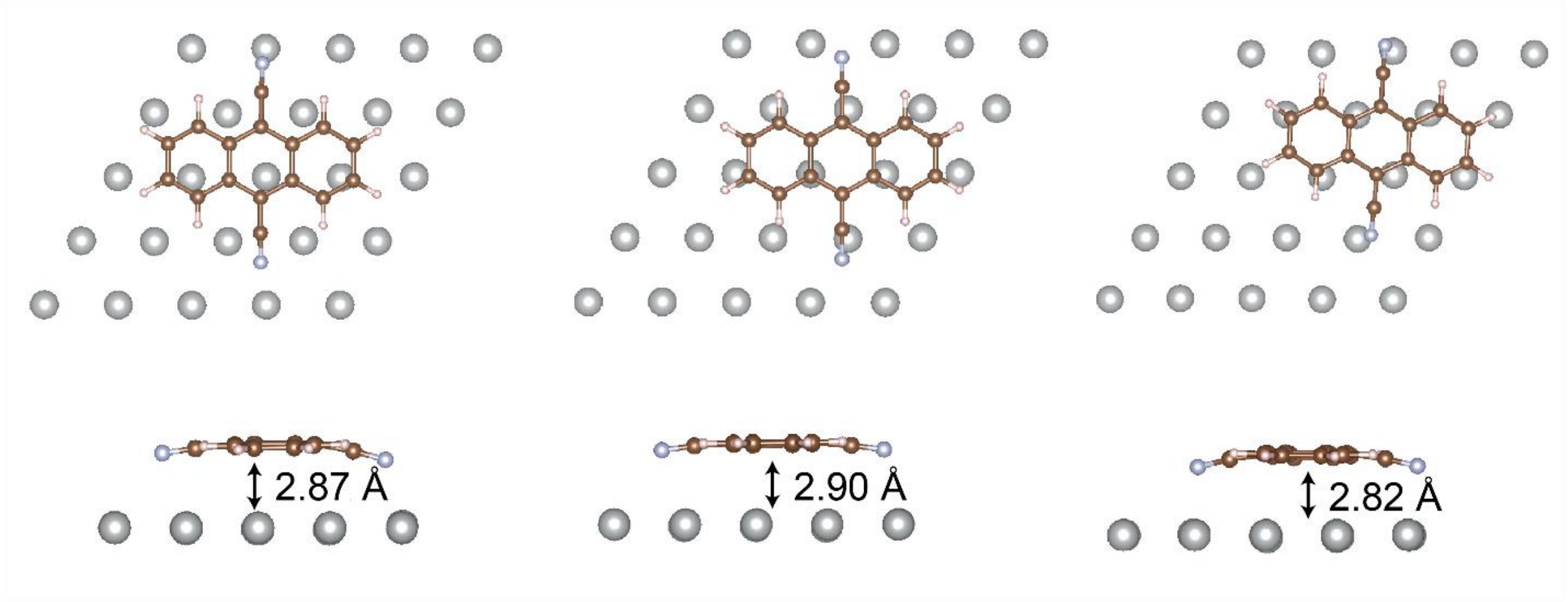

**Figure S9.** DFT-calculated energetically favorable adsorption geometries of DCA on Ag(111), for three inequivalent initial adsorption sites. The shown adsorption heights consist of the average distances between the molecule atoms and the atoms of the Ag surface top layer.

## S9. Formation of DCA dimers at low substrate temperature

By depositing DCA molecules on Ag(111) held at ~5 K, we observed the formation of molecular dimers (Figure S10a). These dimers can be easily moved with the STM tip. This indicates weak molecule-surface interactions, consistent with the observed field-induced charging. We performed $\mathrm{d}I/\mathrm{d}V$ STS on a DCA dimer (red and blue crosses in Figure S10b), showing a prominent dip at bias voltage $V_\mathrm{b} \approx -4.4$ V (Figure S10c; note the dimers seem to be more stable and less prone to bias-induced damage compared to molecules in the nanofilm) and a peak that we associate with the LUMO of the molecule at $V_\mathrm{b} \approx 470$ mV (Figure S10d). We explain the larger $|V_\mathrm{b}|$ (and hence larger electric field magnitude) required to charge the DCA molecule in this dimer case by two factors: (1) the LUMO energy of the DCA molecules in the dimer is significantly higher (~470 mV) with respect to the Ag(111) surface Fermi level, compared to the LUMO energy of the DCA molecules in the self-assembled nanofilm (~300 mV); and (2) greater molecule-surface interaction (resulting in a smaller effective molecule-metal tunneling barrier). The discrepancy in the LUMO energies can be explained by in-plane screening (or lack of in the case of dimers) of surrounding molecules. The LUMO of DCA molecules in the nanofilm is further stabilized (by a larger stabilization energy $\Delta E$; see main text and section S7 above) compared to dimer molecules due to additional screening from surrounding molecules. This leads to a larger electron affinity and lower LUMO energy for molecules in the nanofilm, as seen in other molecular systems.[4] Further, in the dimer case, a greater molecule-surface interaction can be explained by the interaction of the cyano group lone electron pairs with the underlying surface. In the nanofilm case, this molecule-surface interaction is arguably reduced given the participation of the cyano group lone electron pairs in in-plane directional hydrogen bonding with adjacent molecules.

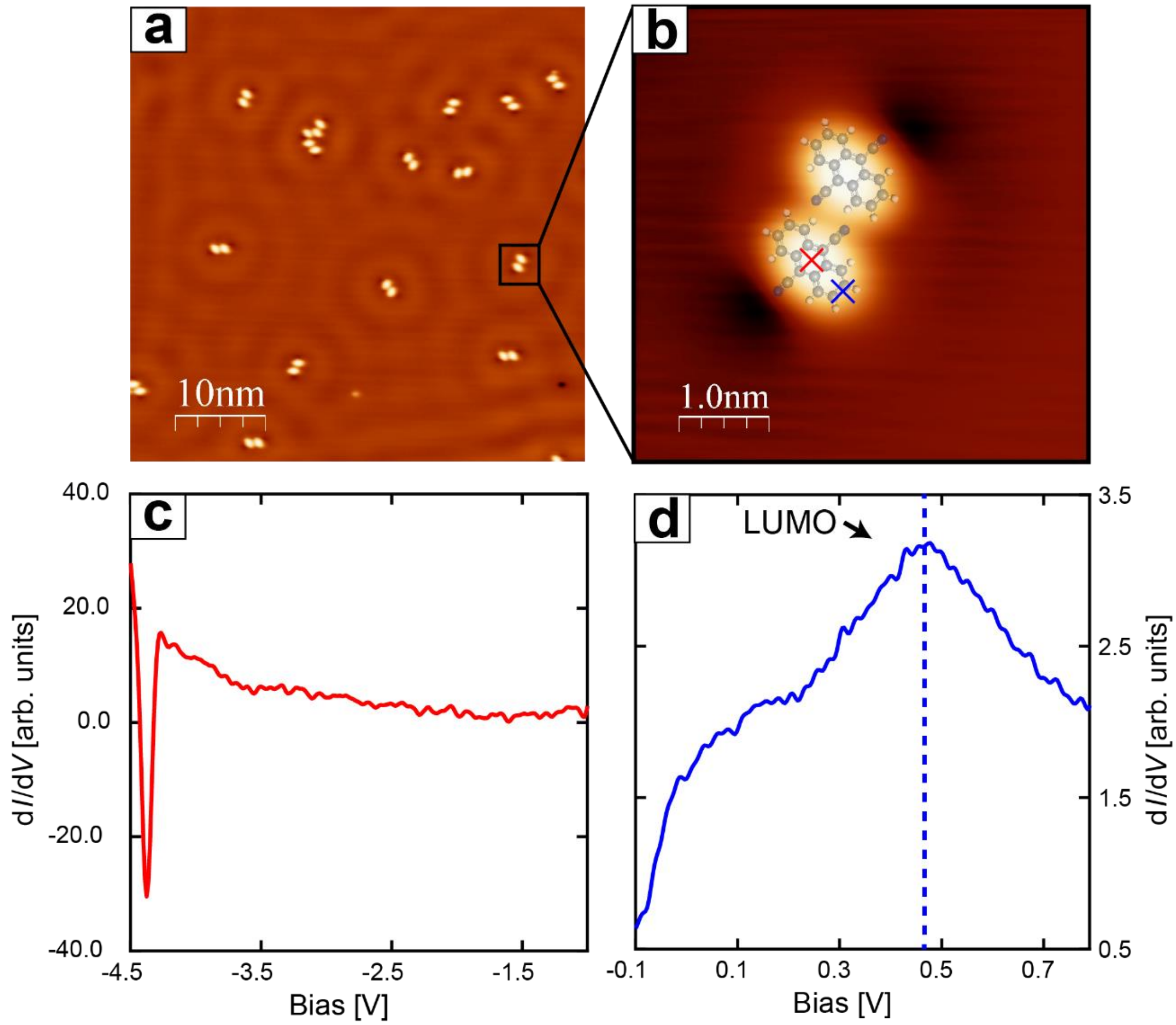

**Figure S10.** (a) Constant-current STM image after deposition of DCA molecules on Ag(111) at 5 K ($V_b$ = -0.02 V, $I_t$ = 200 pA). (b) Zoomed-in STM image of DCA dimer indicated in (a) (black square; $V_b$ = -0.02 V, $I_t$ = 200 pA). (c) – (d) Point d*I*/d*V* STS measurements taken on DCA molecule in (b) in the negative bias range at the centre of the molecule (red cross; set point, $V_b$ = -1.0 V, $I_t$ = 5 pA), and in the positive bias range at the anthracene group extremity (blue cross; set point, $V_b$ = -0.2 V, $I_t$ = 1 pA). The charging dip can be observed at ~-4.4 V and the peak associated with the LUMO (blue dashed line) can be observed at ~0.47 V.

## S10. Local contact potential difference (LCPD) measurements

In order to verify possible local variations of the work functions within a molecular domain, we measured the frequency shift $\Delta f$ (nc-AFM qPlus sensor with Ag-terminated Pt/Ir tip) as a function of bias voltage $V_\mathrm{b}$, for molecules $M_0$ and $M_{20}$ (Figure S11). The bias voltage for which $\Delta f$ is maximum corresponds to the local contact potential difference (LCPD).[8-10] The LCPD for both $M_0$ and $M_{20}$ is $\sim -297$ mV, whereas the charging onset voltage $V^*$ for $M_0$ measured by STS is $\sim -1.6$ V for the same tip-sample distance ($M_{20}$ does not show charging; see main text). This is consistent with the downward kink and deviation of $\Delta f(V_\mathrm{b})$ from a parabolic curve for $M_0$ at $V_\mathrm{b} = -1.66$ V, indicative of bias-induced charging[11-15] (Figure S11). That is, although we observe the effect of field-induced charging of DCA on $\Delta f(V_\mathrm{b})$ at $V_\mathrm{b} = -1.66$ V, it is challenging – if not impossible – to reliably derive the LCPD of negatively charged DCA (*i.e.*, *via* quadratic fitting of $\Delta f(V_\mathrm{b})$ for $V_\mathrm{b} < -1.66$ V) given the large discrepancy between $V^*$ and the bias voltage required to maximize $\Delta f(V_\mathrm{b})$. Measuring the LCPD accurately for negatively charged DCA (and addressing how this LCPD varies as a function of site in a molecular domain) requires decoupling of the charging voltage (*i.e.,* gating) and the LCPD measurement voltage variable. This is beyond the scope of this work and requires further experiments.

Note that retrieving work function changes due to charging *via* $\mathrm{d}I/\mathrm{d}z$ STS (at constant bias voltage) is challenging since varying $z$ results in varying the electric field at the junction, and hence potentially varying the charge state of the molecule. That is, the measurement itself would alter the state of the system.

It is also worth noting that we attempted resolving the intramolecular morphology of negatively charged DCA *via* CO-tip ncAFM imaging.[11, 14] This involves large bias voltage absolute values (required for charging) and small tip-sample distances (required for achieving intramolecular chemical bond resolution and resulting in significant probe-sample interactions). In our case, we observed that these acquisition parameters rendered the CO-functionalized probe less stable, which hindered the measurement.

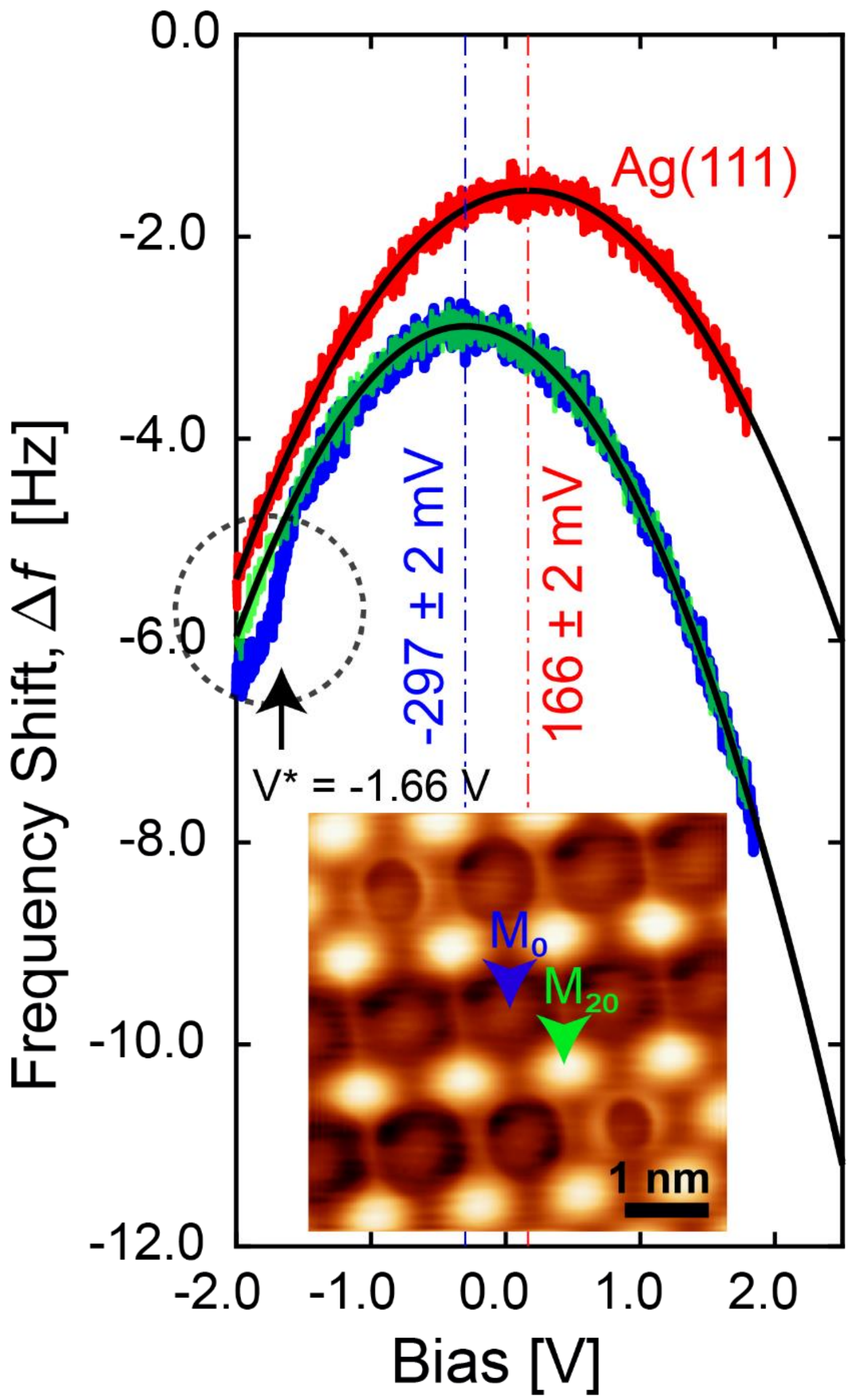

**Figure S11.** Frequency shift $\Delta f$ as a function of bias voltage $V_b$, (set point $V_b$ = –20 mV, $I_t$ = 5 pA on top of $M_0$) measured on molecule $M_0$ (blue curve), $M_{20}$ (green), and on bare Ag(111) (red). The bias voltage for which $\Delta f$ is maximum corresponds to the local contact potential difference (LCPD). Solid black curves are quadratic fits. Inset: constant-current STM image of DCA domain ($V_b$ = –2.4 V, $I_t$ = 50 pA). Obtained LCPD values are -297 ± 2 mV and 166 ± 2 mV for (neutral) DCA and Ag(111), respectively. Kink and downward shift of $\Delta f(V_b)$ (dashed circle) at $V_b$ = -1.66 V (black arrow) for $M_0$ is due to negative charging of DCA. Measurements were performed using an ncAFM qPlus sensor with an Ag-terminated Pt/Ir tip (oscillation amplitude = 100 pm).

## S11. Telegraph current signal of molecule with bistable charge state

Figure S12a shows a constant-current STM image of a DCA domain at a bias voltage $V_\mathrm{b}$ = -2.4 V. At this bias voltage, some molecules are negatively charged (dark molecules) and some are neutral (bright). In between the negatively charged and neutral molecules, along $\vec{a}_1$, we observe molecules that, during the scan, flicker and alternate between dark and bright, that is, between negative and neutral charge state (blue dashed circles in Figure S12a). We observe such molecules with alternating charge state also in constant-height STM imaging ($V_\mathrm{b}$ = -2.0 V; see blue dashed circle in Figure S12b). This is due to the fact that, at the tip-sample distance considered, the charging bias onset $V^*$ of these molecules is close to the bias voltage $V_\mathrm{b}$ at which the STM data were acquired; these molecules are in a bistable charge state. Figure S12d shows the tunneling current (absolute value) as a function of time (derived from the scan speed) during the constant-height scan over the bistable molecule in Figure S12c. We observe discrete changes between low and high tunneling current, equivalent to telegraph current noise, similarly to other bistable atomic and molecular adsorbates.[16] Note that here our signal is convoluted with the topography of the molecule and hence we do not observe transitions between two constant current levels. The bandwidth of our tunneling current line limits our time resolution to ~1 ms; measuring the charge state lifetime accurately would require increased time resolution and is beyond the scope of this work.

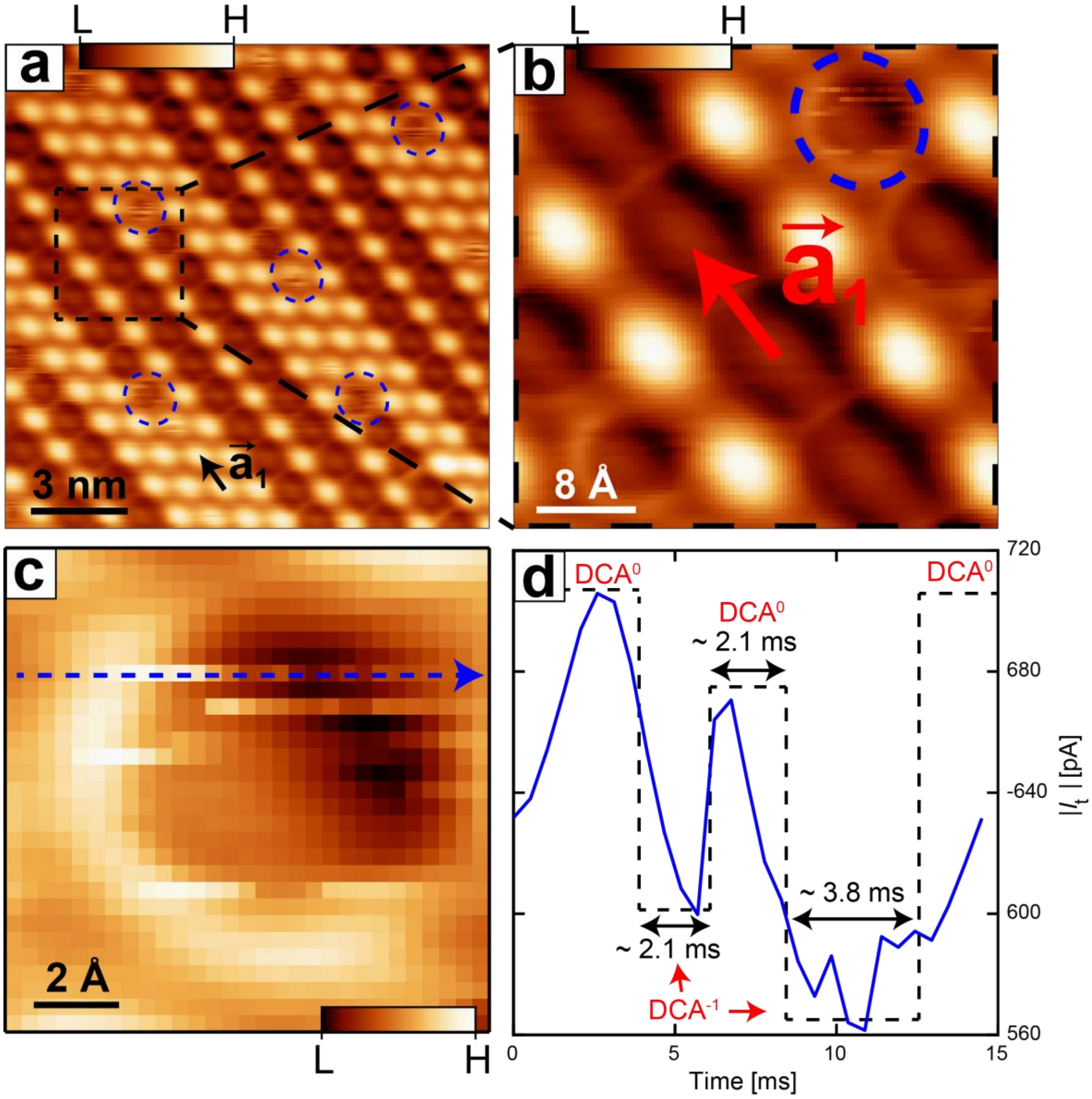

**Figure S12.** (a) Constant-current STM image ($V_b$ = –2.4 V, $I_t$ = 50 pA) of DCA molecular domain. Molecules indicated by blue dashed circles show charge state bistability. (b) Constant-height STM image of dashed square region in (a) (tunneling current absolute value; $V_b$ = –2.0, set point $I_t$ = 50 pA above molecule indicated with blue dashed circle). Blue dashed circle indicates molecule with bistable charge state. (c) Zoomed-in constant-height STM image of molecule indicated by dashed circle in (b). (d) Tunneling current (absolute value) as a function of time during scan along dashed line in (c) over a molecule with bistable charge state. Time variable was derived from tip's scanning speed (78.125 nm/s). Transition from high to low tunneling current (absolute value) corresponds to a change of molecular charge state from neutral ($DCA^0$) to negative ($DCA^{-1}$).

## S12. $dI/dV$ STS measurements for molecules $M_0$ to $M_{39}$ (negative bias range)

Figures S13b, c show point $dI/dV$ spectra for DCA molecules $M_0$ to $M_{39}$ along $\vec{a}_1$ (see Figure S13a) for $-2.7 < V_b < -1.0$ V. Note that, due to the periodicity of this molecular domain, $M_0$ is equivalent to $M_{39}$. We observe that the absolute value $|V^*|$ of the charging bias onset (*i.e.*, bias voltage corresponding to the negative differential conductance dip) increases the further a molecule is from $M_0$ along $\vec{a}_1$ ($M_0$ to $M_{10}$). For molecules $M_{29}$ to $M_{39}$ (*i.e.*, as the distance to $M_0$ diminishes, given that $M_{39}$ is equivalent to $M_0$), $|V^*|$ decreases. Note that these measurements were performed for the same molecules as in Figure 4 of the main text.

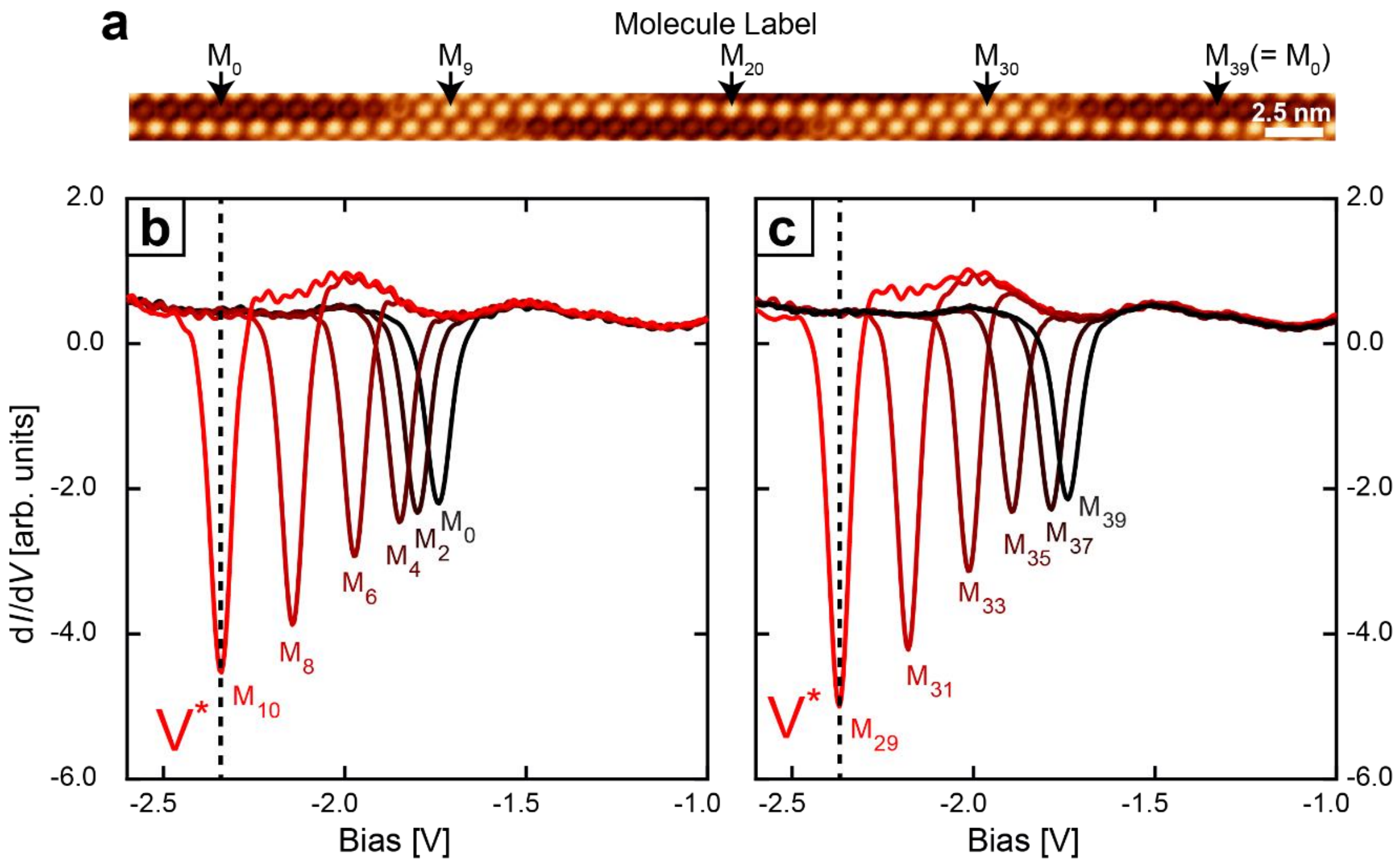

**Figure S13.** (a) Constant-current STM image ($V_b$ = –2.4 V, $I_t$ = 50 pA) of the same molecular domain as in Figure 4 of main text. Here, the bias-dependent charging behavior of DCA along $\vec{a}_1$ repeats after 39 molecules. (b) – (c) $dI/dV$ STS measurements (set point $V_b$ = –20 mV, $I_t$ = 5 pA on top of $M_0$) in the negative bias regime for molecules $M_0$ to $M_{39}$ as indicated in (a). The charging bias onsets, $V^*$, correspond to the voltage positions of the negative differential conductance dips in the spectra.